\documentclass[12pt,reqno]{article}
\usepackage[lmargin=2cm, rmargin=2cm]{geometry}
\usepackage{amsfonts}
\usepackage{amssymb}
\usepackage{enumerate}
\usepackage{amsthm}
\usepackage{array}
\usepackage{epsfig}
\usepackage[small,bf]{caption}
\usepackage{appendix}
\usepackage{dsfont}
\usepackage{hyperref}
\usepackage{amsmath}
\usepackage{graphicx}
\usepackage{epstopdf}
\usepackage{blkarray}
\usepackage{cite}

\allowdisplaybreaks \numberwithin{equation}{section}

\begin{document}

\begin{titlepage}
 \thispagestyle{empty}

\begin{flushright}
 \end{flushright}

 \begin{center}

 \vspace{30mm}

     { \LARGE{\bf  {Black holes, complexity and quantum chaos}}}

     \vspace{40pt}

\Large{{\bf Javier M. Mag\'an}} \\[8mm]
{\small\slshape Instituto Balseiro, Centro At\'omico Bariloche \\
S.~C. de Bariloche, R\'io Negro, R8402AGP, Argentina\\

\vspace{5mm}

{\upshape\ttfamily javier.magan@cab.cnea.gov.ar}\\[3mm]}

\vspace{8mm}

     \vspace{10pt}

    \vspace{10pt}

\date{\today}

\end{center}

\begin{abstract}
We study aspects of black holes and quantum chaos through the behavior of computational costs, which are distance notions in the manifold of unitaries of the theory. To this end, we enlarge Nielsen geometric approach to quantum computation and provide metrics for finite temperature/energy scenarios and CFT's. From the framework, it is clear that costs can grow in two different ways: operator vs `simple' growths. The first type mixes operators associated to different penalties, while the second does not. Important examples of simple growths are those related to symmetry transformations, and we describe the costs of rotations, translations, and boosts. For black holes, this analysis shows how infalling particle costs are controlled by the maximal Lyapunov exponent, and motivates a further bound on the growth of chaos. The analysis also suggests a correspondence between proper energies in the bulk and average `local' scaling dimensions in the boundary. Finally, we describe these complexity features from a dual perspective. Using recent results on SYK we compute a lower bound to the computational cost growth in SYK at infinite temperature. At intermediate times it is controlled by the Lyapunov exponent, while at long times it saturates to a linear growth, as expected from the gravity description.
\end{abstract}

 \vspace{10pt}
\noindent

\end{titlepage}

\thispagestyle{plain}

\tableofcontents

\baselineskip 6 mm

\newpage

\section{Introduction}\label{secI}

Although there is a large amount of knowledge about the holographic dictionary in the context of AdS/CFT \cite{adscft}, see \cite{Harlow} for a recent review on bulk reconstruction and references therein, it remains unclear how the CFT describes processes behind or near the horizon of a black hole. One of the main reasons is that most of the well-known entries of the dictionary consider setups anchored at the boundary of AdS, such as the field operator correspondence \cite{fieldopcorrespondence} or the Ryu-Takayanagi formula for computing entanglement entropy \cite{Takayanagi,LM}. On the other hand, an example of quantities that are transparently sensitive to near horizon dynamics are out of time-ordered correlation functions (OTOC), as developed in \cite{SS,bound}. But these OTOC are sensitive through $\mathcal{O}(1/N)$ effects, both from CFT and gravity points of view, while the near horizon geometry and its physics are $\mathcal{O}(1)$ effects that should be encoded in the CFT through $\mathcal{O}(1)$ effects as well. For example, as we review below, infalling particles have basic properties like energy and momentum that are controlled by the chaos exponent (see for example \cite{Sbook,BM,Sfall}), and one would like to understand how such properties are encoded in the CFT. Besides, as we will see, there are universal features not directly captured by OTOCs.

As in many other physics situations, a great deal of the problem seems to rest on a proper choice of physical quantities. Inspired by the geometric approach to quantum complexity developed by Nielsen and collaborators \cite{Nielsen1,Nielsen2,Nielsen3}, and by the recent ideas which relate gravity and complexity \cite{SusskindQC,Aaronson}, in this article we explore the previous questions with the aid of fine-grained distance notions in the manifold of unitaries (or in the Hilbert space). As with any other distance notion, these will not care so much about the present state of the system, but more about `its entire history'. Important examples of `unitary histories' we will consider are time evolution, generic symmetry transformations, and Heisenberg time evolution. In the field of quantum complexity, these distances are called computational costs, and we will often refer to them in the same way. The intuition behind the name is that unitary evolution defines a protocol (a continuous or `analogic' one, but still a protocol), whose computational cost is just given by a suitable notion of length in the unitary manifold, as we review in detail in the next section.

The article is organized as follows. Section~(\ref{secII}) is devoted to the definition of distances in generic quantum theories. We start by commenting and proposing solutions to some problems in Nielsen geometric approach to quantum computation. We then provide explicit metrics for the manifold of unitaries, valid for finite temperature/energy scenarios and CFT's. We further discuss three aspects which help to clarify and disentangle conceptual and technical problems in this context. The first one comments on the main technical difficulty associated with the actual computation of these distances/costs. This is the problem of finding the infinitesimal unitary transformation that is being applied at each given instant of time, given the knowledge of the unitary trajectory. Such problem has a precise solution, but not an easy one to handle in general. The second one concerns the so-called penalty functions and their role in CFT's. In the field of quantum complexity, penalty functions are included to punish directions which are assumed more difficult to explore. From a geometric perspective, they are just the definition of the local metric, which is to some extent arbitrary. One possibility is that penalties are functions of the operator scaling dimensions in CFT's. The most important argument will be that such choice allows the study of the dynamics of `local' scaling dimensions. These are natural CFT quantities with non-trivial dynamics that, to the author's knowledge, have not been considered in the literature. Finally, we also describe the two qualitatively different ways in which complexity can grow. The first one concerns mixing of operators with different penalties, while the second one does not. Interestingly, this last way, which turns out to be much simpler to compute, is crucial for many applications.

Using the framework and the developed intuition, in section~(\ref{secIV}) we start by considering the behavior of computational costs under symmetry transformations. Symmetry transformations always fall in the class of simple growths. Furthermore, we will show that the technical difficulty can be overcome due to the group structure. In this article, we will study Lorentz transformations (rotations and boosts) and some general coordinate transformations. In a forthcoming article, we will study the conformal group \cite{uspawel}.

After gathering the results and intuition from symmetry transformations, we show, as argued in \cite{Snegative}, that the computational cost of an infalling particle in a black hole increases exponentially with time with the maximum Lyapunov exponent, and therefore directly encodes the near horizon geometry. Interestingly, this statement rests on the equivalence principle. If the particle/system momentum in the freely falling frame is constant, this necessarily implies that the costs described by an outside observer increase exponentially with time. As a byproduct, we show that the present approach suggests a further chaos bound in the coefficient in front of the exponential growth, obtained by letting the infalling particle approach the speed of light. This seems a non-trivial prediction for a dual theory to have a local Minkowskian gravitational dual near the black hole horizon. This should be contrasted to the OTOC approach, for which, to the author's knowledge, the coefficient in front of the exponential growth is operator dependent \cite{bound}.

Finally, in the last section, we study these features of complexity from a microscopic dual perspective. By using recent results on operator growth in SYK \cite{operatorgrowth}, we obtain a lower bound on the computational cost in the dual theory. Before the scrambling time, we confirm it is controlled by the Lyapunov exponent. After the scrambling time, the exponential growth saturates to a linear growth. On one hand, this dramatic change in the dynamics is actually mirrored in the gravitational description, since by times of the order of the scrambling time backreaction of the infalling shock wave has to be included, due to its large proper energy \cite{SStanford}. On the other hand, this dynamical transition is consistent with Lloyd's bound \cite{lloyd}, since an indefinite exponential growth would, after the scrambling time, completely invalidate it.

Before we move on, we want to make a couple of general comments. First, notice that well-defined distance notions in the Hilbert space or in the manifold of unitaries have to be respected across dualities, so the present approach is self-consistent. An important example in holography is relative entropy. In \cite{JLMS}, it was proven that bulk relative entropy equals CFT relative entropy, as it should, given the assumed equality of Hilbert spaces. The problem is that bulk relative entropy resists a meaningful definition since it is related to the vacuum entanglement of quantum fields in the bulk. Another problem is that it is anchored in the boundary, complicating the exploration of the full geometry and the exportation of the technique to more general spacetimes. Finally, when considering pure state scenarios, relative entropy, being invariant under unitary transformations, is not fine grained enough to certain details of the evolution. Computational costs seem to avoid all of these problems. They are well defined and computable at both sides of the duality, they are able to explore the full geodesic structure, the method is not attached to any particular geometry, and they are perfectly suited for pure state contexts.

Second, one possible reason why these type of interesting quantities have passed largely unnoticed in the physics community is because usual statistical ensembles, which can be used to build well defined distance notions, are totally blind to details of time evolution. For example:
\begin{equation}
\textrm{Tr}(\,\rho_{\beta}\,U(t)^{-1}\,V_{0}\,U(t)\,)=\textrm{constant}\:,
\end{equation}
or
\begin{equation}
\textrm{Tr}(\,\rho_{\beta}\,\frac{dV(t)^{\dagger}}{dt}\,\frac{dV(t)}{dt}\,)=\textrm{constant}\:.
\end{equation}
As we will see, when choosing the standard metric in the unitary manifold, the last expression is what needs to be evaluated to find distances. It is clear that finer grained notions of length are needed, and suitable notions can be found by generalizing Nielsen geometric approach to quantum computation \cite{Nielsen1,Nielsen2,Nielsen3}. For recent related approaches to quantum complexity in physics see \cite{Myers,Chapman,Yang,PawelC,Norihiro}.

\newpage

\section{A geometric approach to quantum mechanics}\label{secII}

As described in the introduction, in the context of dualities it is interesting to have fine-grained notions of distance in the unitary manifold, since these have to be preserved across the duality. Typical interesting distances are those associated with Hamiltonian time evolution:
\begin{equation}
U(s)=e^{-iHs}\;,
\end{equation}
and any type of symmetry transformations:
\begin{equation}
U(s)=e^{-i\sum\limits_{j}\theta_{j}(s)T_{j}}\;,
\end{equation}
associated to some set of Lie algebra generators $T_{j}$ of certain symmetry group $G$. Finally, we can be interested in the cost of  Heinsenberg time evolution:
\begin{equation}
V(t)=U(t)^{-1}\,V\,U(t)\;,
\end{equation}
where $V$ is a unitary perturbation of the state.

Mathematically, the problem is to assign lengths to trajectories $U(s)$ in the unitary manifold. Generically, lengths are defined by integrating a suitable `norm' of the tangent vector to the trajectory along the curve. For Riemannian geometry, the famous expression reads:
\begin{equation}
L(x(s))=\int\limits_{s}ds\,\sqrt{g_{\mu\nu}\frac{dx^{\mu}}{ds}\frac{dx^{\nu}}{ds}}\;,
\end{equation}
so what we need is a chart $x^{\mu}$ and a metric on the tangent space.

In what follows we follow Nielsen approach \cite{Nielsen1,Nielsen2,Nielsen3} to define the geometry. The motivations to define such geometries were purely related to quantum computation. The objective was to define geometries such that the lengths of minimal geodesics provide lower bounds to quantum complexity. It is perfectly possible that for physics applications there exist other metric definitions that are also sufficiently fine grained. In this article, we will concentrate and expand on these quantum complexity inspired notions of length, which will suffice for our purposes. But at any rate, we first want to remark that such geometric approach is just the natural mathematical approach to define distances in the unitary manifold, and also we want to remind that for applications in dualities, the only important thing is that we use the same geometry on both sides of the duality and that the geometry is sufficiently fine-grained.

The starting point is that the added computational cost (the added distance in the geometry) that arises when applying a small unitary evolution to any unitary matrix is independent of the input. In other words, for the infinitesimal gate/transformation:
\begin{equation}\label{left}
U(s+ds)=e^{-i\tilde{H}(s)\,ds}U(s)\:,
\end{equation}
the added distance does not depend on $U(s)$, and it is just a function of the instantaneous Hamiltonian $\tilde{H}(s)$ being applied at time $s$ to move us from $U(s)$ to $U(s+ds)$. This simple condition is just a short of local flatness in the unitary manifold. It is just the way to impose that the cost of applying a gate is an intrinsic property of the gate itself. We use the tilde $\tilde{H}(s)$ notation to distinguish the instantaneous Hamiltonian $\tilde{H}(s)$ from the Hamiltonian of the physical theory $H$ since generically they will be totally different objects.

So if we are interested in analyzing certain unitary history $U(s)$, we are forced to find $\tilde{H}(s)$, such that~(\ref{left}) holds at each point of the trajectory. This instantaneous Hamiltonian $\tilde{H}(s)$ turns out to be given by the associated Schrodinger equation:
\begin{equation}\label{Hs}
\tilde{H}(s)=i\,\frac{d U(s)}{ds}\,U^{\dagger}(s)\;.
\end{equation}
From a physical perspective, the instantaneous Hamiltonians $\tilde{H}(s)$ provide the `velocities' used to explore the unitary manifold. They are elements of the tangent space. This situation exactly parallels the analysis of Lie groups, as we will exploit later in the article. Notice that the previous relation holds because at first order in $ds$ we have:
\begin{equation}
U(s+ds)=U(s)+\frac{d U(s)}{ds}\,ds = e^{-i\,(i\frac{d U}{ds}U^{\dagger}) \,ds}\,U(s)\;.
\end{equation}
Once we find $\tilde{H}(s)$ from $U(s)$, the computatinal cost associated to the trajectory $U$ is given by its length:
\begin{equation}
C(U)\equiv L(U)=\int F(\tilde{H}(s))\, ds\;,
\end{equation}
where $F(\tilde{H}(s))$ is a metric functional on the tangent space of the unitary manifold. Before defining $F$, let us remark that computing $\tilde{H}(s)$ from~(\ref{Hs}) can be quite non-trivial, as we explain in detail in the next section. The simplest examples are those in which the unitary evolution can be written as:
\begin{equation}
e^{-i H s}\;,
\end{equation}
for which $\tilde{H}(s)=H$. In these cases $C(U)=F(H)s$, and this is how the famous linear growth of complexity looks like in the geometric approach.

Let us continue and define the metric $F$. We first need to define a chart and this will depend on the theory. We will consider explicit examples later, but for the time being, we just assume there is an orthonormal basis of hermitian generators  $T_{\mu i}$ of the tangent space. This allows us to write all Hamiltonians as:
\begin{equation}\label{H}
\tilde{H}(s)=\sum\limits_{\mu i}x^{\mu i}(s)T_{\mu i}\;.
\end{equation}
We include two indices because one will run over operators with different penalty factors (index $\mu$), and the other over operators associated to the same penalty (index $i$). Of course, in explicit examples like SYK or CFT's, the sum over $i$ might implicitly depend on the specific $\mu$.

For the present dicussion we assume the manifold is finite dimensional. We will later generalize the framework to CFT's. So if we are given $\tilde{H}(s)$, and the dimension of the Hilbert space is $\vert\mathcal{H}\vert$, the expansion coefficients are given by:
\begin{equation}
x^{\mu i}(s)=\frac{1}{\vert\mathcal{H}\vert}\textrm{Tr}\,(H(s)T_{\mu i}\,)\;.
\end{equation}
In this generic context, the (sufficiently fine grained) class of metrics proposed in \cite{Nielsen1} is:
\begin{equation}\label{metric}
F(\tilde{H}(s))=\sqrt{\sum\limits_{\mu i}(x^{\mu i}(s)p_{\mu})^{2}}=\sqrt{\sum\limits_{\mu}(\sum\limits_{i}x^{\mu i}(s)^{2}) p_{\mu}^{2}}\;,
\end{equation}
where $p_{\mu}$ are some unkown penalty factors that are included to differentiate between various directions in the manifold. We defined them as $p_{\mu}^{2}$ so that when we are evolving in only one direction, associated to one specific generator $T_{\mu i}$, the cost is proportional to $p_{\mu}$. In the context of computational complexity, the penalties $p_{\mu}$ are included to punish directions associated to operators that are assumed to be more difficult to apply (or create), but no generic principle is given to find them. We will comment on them below.

For later reference, notice that it is natural to define the projector into the space of generators with equal penalty factors $p_{\mu}$:
\begin{equation}
\hat{P}_{\mu}(\tilde{H}(s))=\sum\limits_{i}x^{\mu i}T_{\mu i}\;,
\end{equation}
where there is no summation over the index $\mu$. Using such projector, the metric~(\ref{metric}) can also be written in two suggestive ways:
\begin{equation}\label{ex1}
F_{1}(\tilde{H}(s))= \sqrt{\,\sum\limits_{\mu}p_{\mu}^{2}\,\textrm{Tr}(\,\rho_{\textrm{mixed}} \,P_{\mu}(\tilde{H}(s))P_{\mu}(\tilde{H}(s))\,)}\;,
\end{equation}
and
\begin{equation}\label{ex2}
F_{2}(\tilde{H}(s))= \sqrt{\,\sum\limits_{\mu}p_{\mu}^{2}\,\textrm{Tr}(\,\rho_{\textrm{mixed}} \,\tilde{H}(s)\,P_{\mu}(\tilde{H}(s))\,)}\;,
\end{equation}
where $\rho_{\textrm{mixed}}=\mathds{1}/\vert H\vert$ is the usual maximally mixed density matrix. From this perspective, the reason to have chosen $\rho_{\textrm{mixed}}$ and not any other state is unjustified. One interpretation is that the cost in a given state is given by the previous relations, but with the new state inserted in the position of $\rho_{\textrm{mixed}}$. Then, if we have some average over states, such as $\rho_{\textrm{mixed}}$, the associated cost is just the average of the costs. This comment will become clearer when generalizing to QFT's below. Notice that the reason why the previous two expressions are equal is that correlations between generators associated to different penalties vanish.

In the next sections, we enlarge this framework so as to include manifolds with infinite dimensions (like CFT's), situations in which we are at finite temperature or energy, and comment on the issue of penalty factors. We will also discuss the technical difficulties that appear in actual computations, and make some simple but important remarks on the possible types of complexity growth we might have.

\subsection{State dependence}\label{secII.I}

The past framework was fairly generic, and it exactly parallels Nielsen's approach to finite dimensional spin systems \cite{Nielsen1}. But there are a couple of issues that need to be faced in order to go towards physics applications. The first concerns the extension to infinite dimensional systems like QFT's, or even to finite dimensional systems but at finite temperature.

For example, consider the Hamiltonian of a free QFT:
\begin{equation}
H_{\textrm{QFT}}=\sum\limits_{k}\omega_{k}a^{\dagger}_{k}a_{k}\;.
\end{equation}
Blindy using~(\ref{metric}), the cost of such an operator would be:
\begin{equation}\label{costQFT}
F(H_{\textrm{QFT}})=\sqrt{\sum\limits_{k}(\omega_{k}p_{a^{\dagger}_{k}a_{k}})^{2}}\;,
\end{equation}
where $p_{a^{\dagger}_{k}a_{k}}$ is the penalty associated to the number operator $a^{\dagger}_{k}a_{k}$. There are two problems with~(\ref{costQFT}). The first is that, unless $p_{a^{\dagger}_{k}a_{k}}$ decays sufficiently fast with $k$, which seems totally unphysical, the answer diverges. The second is that, even in the unphysical case in which the answer is finite, in a real situation we would be counting the cost of operators that are not being used. For example, if we are in a state $\vert\psi_{k_{\textrm{max}}}\rangle$ in which there are no particles with momenta higher than $k_{\textrm{max}}$, the action of $H_{\textrm{QFT}}$ on the state is equal to the action of:
\begin{equation}
H_{\textrm{QFT}}^{k_{\textrm{max}}}=\sum\limits_{k}^{k_{\textrm{max}}}\omega_{k}a^{\dagger}_{k}a_{k}\,
\end{equation}
whose cost, using again formula~(\ref{metric}), is finite and given by:
\begin{equation}
H_{\textrm{QFT}}^{k_{\textrm{max}}}=\sqrt{\sum\limits_{k}^{k_{\textrm{max}}}(\omega_{k}p_{a^{\dagger}_{k}a_{k}})^{2}}\,
\end{equation}
The moral is straightforward. Since the action of the high momentum Hamiltonian tail on the state is equal to zero, and zero has vanishing cost, we would like to say that the cost of $H_{\textrm{QFT}}$ in the state $\vert\psi_{k_{\textrm{max}}}\rangle$ is equal to the cost of $H_{\textrm{QFT}}^{k_{\textrm{max}}}$. At first sight, this might appear like some short of state dependence, but it is actually not. We are just choosing the operator that minimizes complexity costs, while still moving in the same trajectory of the Hilbert space. Besides, the cost does not depend on a putative previous unitary trayectory, so it is still an intrinsic property of the gate itself.

We need to formalize this intuition so as to be applicable to generic situations. The first option is to consider the expectation value of the Hamiltonian in the present state:
\begin{equation}
F(H_{\textrm{QFT}},\vert\psi_{k_{\textrm{max}}}\rangle)=\sqrt{\langle\psi_{k_{\textrm{max}}}\vert H_{\textrm{QFT}}^{\dagger}H_{\textrm{QFT}}\vert\psi_{k_{\textrm{max}}}\rangle}=\sqrt{\sum\limits_{k}^{k_{\textrm{max}}}(\omega_{k})^{2}}\;.
\end{equation}
This gives a finite answer, but it does not capture a possible dependence on the penalties. A simple route that does capture the penalty dependence goes by looking at the previous alternative metric formulation~(\ref{ex1}). The natural generalizations for the cost of $\tilde{H}(s)$ in the state $\vert\psi\rangle$ are:
\begin{eqnarray}\label{metricstate1}
&F_{1}(\tilde{H}(s),\vert\psi\rangle)\equiv & \sqrt{\,\sum\limits_{\mu}\,p_{\mu}^{2}\,\langle\psi\vert P_{\mu}(\tilde{H}(s)^{\dagger})P_{\mu}(\tilde{H}(s))\,\vert\psi\rangle}\nonumber\\
&F_{2}(\tilde{H}(s),\vert\psi\rangle)\equiv& \sqrt{\,\sum\limits_{\mu}\,p_{\mu}^{2}\,\vert\langle\psi\vert \tilde{H}(s)P_{\mu}(\tilde{H}(s))\,\vert\psi\rangle\vert}\;.
\end{eqnarray}
A couple of important remarks are in turn. First, in the generic case the two different choices are not equal. The reason they were equal before is that correlations between different penalty generators in the maximally mixed state vanish, but this is not neccessarily true in generic states. This is a subtle issue because it does not show up easily. Indeed, for the computations below and the ones in the forthcoming article \cite{uspawel}, both definitions are equal.

The second remark is that it seems this definition fails when applied to Hamiltonian eigenstates. For such states, we expect the complexity to not increase. On the other hand, blindly applying the previous relationship with the true Hamiltonian of the system seems to have some cost. The error lies in that in such scenario, we should not insert the Hamiltonian but the identity operator ($\tilde{H}(s)=0$), which has zero cost. Indeed, the general correct way would be to quotient out by the subgroup of unitaries that leaves invariant the given state, and as the representative of a given class of instantaneous Hamiltonians, choose the one that minimizes the previous relation. This is an utterly cumbersome and unpractical definition. Luckily, for physical applications, and in particular applications to chaotic systems, these subtleties do not really matter, since the action of any non-trivial gate will produce some non-trivial change on the state. Moreover, for AdS/CFT applications, it is convenient to have a formula that tells us that the complexity of an energy eigenstate grows linearly with time, since most probes of the state will not distinguish between an eigenstate or a non-equilibrium unitarily evolving state. Finally, the previous distance is the correct notion of distance in the manifold of unitaries, the only subtleties arising when interpreting it as a distance in the Hilbert space.

\subsection{On penalty functions and CFT's}\label{secII.II}

The second problem concerns the weights $p_{\mu}$. In Nielsen's approach to quantum spin systems, the proposed penalties are functionals of the so-called `weight' of the generalized spin operator. A generalized spin operator has the following form:
\begin{equation}
\sigma=\sigma_{x}\otimes\sigma_{y}\otimes\mathds{1}\otimes\cdots\otimes\sigma_{x}\;.
\end{equation}
If there are $N$ tensor product factors, out of which $M$ factors are equal to $\mathds{1}$, then the weight $w$ is equal to $N-M$. The proposed penalties $p_{\sigma}$ in spin systems are functionals of the weight $p_{\sigma}=p_{\sigma}(w)$, that increase as the weight increases. This is how we punish directions that are suppossed to be more `complicated' than others.

In this context, there are a couple of questions that need to be solved for physics applications. The first asks for a more unique functional $p_{\sigma}(w)$ for spin systems, some sort of `natural' penalty functions. The second asks about the role of weight $w$ in generic theories, in particular, their role in CFT's. Concerning the first question, we are going to leave it open for the time being. We will come back to it below~(\ref{secII.II.I}) and in the last section. In both sections, due to different reasons and by exploring different possibilities, we will argue that a good physical choice is $p_{\sigma}(w)=w$.

For the second question, we want to explore the proposal that in CFT's, the role of weight $w$ is played by the scaling dimension $\Delta$ of the associated operator $\mathcal{O}_{\Delta}$. In CFT's, due to the operator product expansion, any gate or any instantaneous Hamiltonian can be expanded in terms of local operators at some fixed time slice:
\begin{equation}\label{hcft}
\tilde{H}=\sum\limits_{\Delta,l}\int d\Sigma^{d-1} \,c_{\mathcal{O}_{\Delta,l}}(x)\,\mathcal{O}_{\Delta,l}(t,x)\;,
\end{equation}
where $x$ is a generalized coordinate for the $d-1$ dimensional spacelike surface at time $t$, and the sum runs over primaries and descendants as well. Another possibility would be to think in radial quantization and expand in terms of operators at some fixed radial slice.

Importantly, notice that we do not need operator products as in the spin system. Due to the OPE and the operator state correspondence, operator growth in CFT's is equivalent to the usual evolution of a quantum state, where the initial state (operator) gets mixed with other states (operators) as time evolves.

As for the the finite dimensional case, once we have characterized the set of infinitesimal gates, we need to define a norm on them. Through the penalties, this norm will tell us which directions in~(\ref{hcft}) are more difficult to explore. Now, given translation invariance, the penalties cannot depend on $x$. They are therefore intrinsic functions of $\mathcal{O}_{\Delta, l}$. This suggests that penalties only depend on the scaling dimension of the operator. It is then natural to define pojectors into subsapces of equal scaling dimension:
\begin{equation}
\hat{P}_{\Delta}(\tilde{H})\equiv \sum\limits_{\mathcal{O}^{i}_{\Delta,l}}\int d\Sigma^{d-1}\, c_{\mathcal{O}^{i}_{\Delta,l}}(x)\,\mathcal{O}^{i}_{\Delta,l}(t,x)\;,
\end{equation}
where the sum runs over all operators with scaling dimension $\Delta$ (again primaries and descendants as well). The generalization of the previous metrics to CFT's is:
\begin{eqnarray}\label{metricstate2}
F_{1}(\tilde{H},\vert\psi\rangle)&\equiv & \sqrt{\,\sum\limits_{\Delta}p (\Delta)^{2}\,\langle\psi\vert\hat{P}_{\Delta}(\tilde{H})^{\dagger}\hat{P}_{\Delta}(\tilde{H})\vert\psi\rangle}\nonumber\\
F_{2}(\tilde{H},\vert\psi\rangle)&\equiv &\sqrt{\,\sum\limits_{\Delta}p (\Delta)^{2}\,\langle\psi\vert (\tilde{H})^{\dagger}\hat{P}_{\Delta}(\tilde{H})\vert\psi\rangle}\;.
\end{eqnarray}
In situations in which there is an approximate continuous spectrum of scaling operators, one can approximate the the sums by continuous integrals, weighted by the degeneracy of the sector of scaling dimension $\Delta$. Notice also that the integrand in the previous expression is finite and positive definite.

There are various reasons suggesting that $\Delta$ plays the role of $w$ in CFT's. First, notice that translational invariance, together with the form of the expansion~(\ref{hcft}), implies that the penalty function must be an intrinsic property of the operator. There are not too many options. The real dimension of the operator/field is not a good choice since the dimension could be zero. Good examples of this situation are the fermions of SYK. In that situation, any string of operators, no matter how large, would still have dimension zero. Another possibility is to count the number of operators in a string of operators, completely paralleling the idea of weight in spin systems. The first problem with this is that, in the context of AdS/CFT, we would be equally punishing an operator with low scaling dimension, that creates a perturbative particle in AdS, with an operator with very large scaling dimension, that is dual to a (pure state) black hole. This seems unreasonable. The second problem with this option is that it is not consistent with the OPE, since any such string can be written as a linear combination of operators with no products whatsoever.

On the other hand, the scaling dimension of the operator seems the right intrinsic property that tells us what is more difficult/easy to create in CFT's. The first reason is that in CFT's, the scaling dimension in radial quantization in the plane turns out to be the energy in the cylinder formulation. Given that complexity generically evolves as $C=E\,t$, where $E$ is the energy of the state, states with higher scaling dimension have correspondingly higher complexity rate growth. If a given state is able to produce more complexity, it should be more difficult to create and should be punished accordingly. In CFT's, this translates into a dependence on the scaling dimensions. In the same line of thought, the Hilbert space of a conformal family is very much like a harmonic oscillator. To obtain a descendant of level $n$ with scaling dimension $\Delta_{n}=\Delta+n$, we need to apply $n$ times the momentum operator. If penalties are functionals of such number $n$, then they are functionals of the associated $\Delta_{n}$. More generically, it seems that the scaling dimension would be a convenient choice when studying how complexity behaves under conformal transformations. In particular, it behaves selfconsistently under the operator product expansion. It is obvious that the penalty could also depend on the operator spin $l$, but we have not found a good specific use for this. Including such dependence is straightforward and carries no conceptual problems. 

Another argument goes by looking at SYK (to be defined below), which is both a spin system and a CFT. In SYK, the scaling dimension of a string of operators in the large-N limit is directly proportional to the number of Majorana fermions in the string. In this case, the scaling dimension proposal reduces to the usual weight prescription. 

Finally, the proposals~(\ref{metricstate2}) will be used in \cite{uspawel}, when studying the complexity of the Virasoro group and CFT's in 1+1 dimension. In such scenarios they will lead to a direct gravitational interpretation.

\subsubsection{Average scaling dimension and natural penalty factors}\label{secII.II.I}

From a physicist point of view, the previous unknown penalty functions $p(\Delta)$ are quite disturbing. There is obviously too much freedom. One expectation is that most choices give similar qualitative results, albeit with certain quantitative differences. This is actually what we will find below for a big class of choices. But at any rate, we would like to have some definite option.

Below, when computing complexity in SYK and comparing with chaos, we will conclude that a good prescription for local fields is given by:
\begin{equation}
p(\Delta)=\Delta\;,
\end{equation}
Actually, since complexity~(\ref{metricstate2}) is defined up to a global choice of units, it is convenient to divide all penalties by the one of the real Hamiltonian of the system (the penalty associated to the energy-momentum tensor). This choice of units obviously ensures that the Hamiltonian has an associated penalty equal to $1$, and the complexity of unitary evolution is simply set to $C(e^{-iH t})=Et$.

Now, has the choice $p(\Delta)=\Delta$ some physical explanation? Is there a natural quantity that carries information about the penalty functions? Here we will argue that there is one indeed.

The previous expansion of the instantaneous Hamiltonian~(\ref{hcft}), together with the projectors into spaces of equal scaling dimension, naturally defines the following probability distributions:
\begin{equation}
\textrm{Prob}_{1} (\Delta)=\frac{\langle\psi\vert\hat{P}_{\Delta}(\tilde{H})^{\dagger}\hat{P}_{\Delta}(\tilde{H})\vert\psi\rangle}{\sum\limits_{\Delta}\textrm{Prob}_{1}(\Delta)}\equiv\frac{\langle\psi\vert\hat{P}_{\Delta}(\tilde{H})^{\dagger}\hat{P}_{\Delta}(\tilde{H})\vert\psi\rangle}{Z_{1}}\;,
\end{equation}
for the first defintion and
\begin{equation}
\textrm{Prob}_{2} (\Delta)=\frac{\vert\langle\psi\vert\tilde{H}\hat{P}_{\Delta}(\tilde{H})\vert\psi\rangle \vert}{\sum\limits_{\Delta}\textrm{Prob}_{2}(\Delta)}\equiv\frac{\vert\langle\psi\vert\tilde{H}\hat{P}_{\Delta}(\tilde{H})\vert\psi\rangle \vert}{Z_{2}}\;,
\end{equation}
for the second.

They can be interpreted as the probability that the operator has dimension $\Delta$. The intuition is that we look at the expansion~(\ref{hcft}) as a state that is expanded on a certain basis of states, and we are defining the probability of finding a state with scaling dimension $\Delta$. 

Having such probability distributions, we naturally look for the average scaling dimension:
\begin{equation}\label{aver}
\overline{\Delta}_{i}\equiv\sum\limits_{\Delta}\Delta \,\textrm{Prob}_{i}(\Delta)\;.
\end{equation}
We see that if we want to compute the average scaling dimension~(\ref{aver}), this is conceptually similar to the computation of the cost~(\ref{metricstate2}), if the penalty functions are set to:
\begin{equation}
p(\Delta)=\Delta\;.
\end{equation}
With this choice, the CFT metrics~(\ref{metricstate2}) take the following natural form:
\begin{equation}
F_{i}(\tilde{H},\vert\psi\rangle)\equiv \sqrt{\,Z_{i}\sum\limits_{\Delta}\Delta^{2}\,\textrm{Prob}_{i}(\Delta)}=\sqrt{Z_{i}\,\overline{\Delta^{2}}(t)}\;.
\end{equation}
This is just the average of the square of the scaling dimension, a natural quantity as well. Besides, whenever the probability distribution is peaked around some definite scaling dimension and normalizing the complexity by $Z_{i}$ the cost is just given by the average scaling dimension defined before. We will consider such objects in SYK below to clarify in a explicit example the differences and similarities. 

\subsection{Technical difficulties with geometric complexity}\label{secII.t}

In the previous sections, we have defined geometries for the manifold of unitaries of generic quantum theories. In principle, such information is enough to compute the lengths of any given trajectory $U(s)$. In practice, as noted in \cite{Nielsen1}, there is a technical obstruction which enormously complicates the problem. In this section, we want to present such technicality, since in the applications below we will have to deal with it.

To have a specific situation in mind, consider we want to compute the cost of Heisenberg time evolution, which is the the length of the following orbit:
\begin{equation}\label{Hein}
U(\mathcal{O},t)\equiv e^{iHt}e^{i\mathcal{O}}e^{-iHt}=e^{i\mathcal{O}(t)}\;,
\end{equation}
where $\mathcal{O}(t)\equiv e^{iHt}\mathcal{O}e^{-iHt}$. To compute the cost, we first need to extract the instantaneous Hamiltonian $\tilde{H}(t)$ that it is being applied at each differential amount of time along the time evolution. This was derived in the previous section to be:
\begin{equation}
\tilde{H}(t)=i\,\frac{d U(\mathcal{O},t)}{dt}\,U(\mathcal{O},t)^{\dagger}\;.
\end{equation}
Quite surprisingly, this simple looking equation is difficult to handle in general, even having the exact $\mathcal{O}(t)$. The reason can be seen as follows. At linear order in $dt$, we can write the following equation:
\begin{equation}
U(t+dt)=e^{-i\tilde{H}(t)dt}e^{-i \mathcal{O}(t)}=e^{-i (\mathcal{O}(t)+dt\frac{d\mathcal{O}(t)}{dt})}\;,
\end{equation}
Given such relation, the instantaneous $\tilde{H}(t)$ can be found in terms of $\mathcal{O}(t)$ and $\frac{d\mathcal{O}(t)}{dt}$ by means of a version of the Baker-Campbell-Hausdorff formula:
\begin{equation}\label{HXY}
\tilde{H}(\mathcal{O}(t),\frac{d\mathcal{O}(t)}{dt})=i \textrm{ad}_{\mathcal{O}(t)}^{-1}(e^{-i\textrm{ad}_{\mathcal{O}(t)}}-\mathds{1}) (\frac{d\mathcal{O}(t)}{dt})=\sum\limits_{j=0}^{\infty}\frac{(-i\textrm{ad}_{\mathcal{O}(t)})^{j}}{(j+1)!}(\frac{d\mathcal{O}(t)}{dt})\;,
\end{equation}
where $\textrm{ad}_{\mathcal{O}(t)}(\frac{d\mathcal{O}(t)}{dt})=[\mathcal{O}(t),\frac{d\mathcal{O}(t)}{dt}]$. So to compute the cost for a given $\mathcal{O}(t)$, we need to evaluate~(\ref{HXY}) and insert it into the metric~(\ref{metricstate1}). Given the previous chains of nested commutators, this certainly seems a challenging task. Below we will see how such task can be accoomplished when the unitary trajectory belongs to some symmetry group. In such cases, the group structure allows ressumation of the series.

\subsection{Simple growths vs operator growths}\label{secII.III}

To end with all these preliminaries, we want to make an important simple remark. There are two qualitatively different ways in which the computational costs can increase along the unitary trajectory.  Consider that the initial Hamiltonian is given by one particular generator, say $\tilde{H}(0)=T_{\nu j}$ \footnote{This generator could just be the momentum operator, a smeared field in a CFT or a fermion in SYK.}. We could have a situation in which the initial generator continues to be the instantaneous Hamiltonian at all points in the trajectory:
\begin{equation}
\tilde{H}(s)=\sum\limits_{\mu i}x^{\mu i}(s)T_{\mu i}=x^{\nu i}(s)T_{\nu i}\;,
\end{equation}
where there is no summation in the last expression. Plugging such formula in any of the metrics defined in the previous sections, we observe that the cost is given by:
\begin{equation}\label{simple}
L(U)=\int F(\tilde{H}(s))\, ds= p_{\nu}\int ds \sqrt{\langle\psi\vert \tilde{H}(s)^{\dagger}\tilde{H}(s)\vert\psi\rangle}\;.
\end{equation}
We will call such cases `simple growths' since they do not imply a mixing of the initial generator with other generators as we proceed along the unitary trajectory. These cases are simpler because the time dependence is all encoded in the intensity change $x^{\nu i}(s)$, which is a fairly common expectation value. The specific penalty factors are not important in order to understand the dynamics. They just factor out. An explicit example is $e^{-iHt}$, whose cost is given by $C(U)= E t$. But there are other non-trivial situations of this simple short, as we show below. A related situation is one in which the initial generator gets mixed with other generators, but only with those with the same penalty factor $p_{\nu}$. In such case, the cost expression~(\ref{simple}) still holds. These simple cases appear naturally when considering unitary paths generated by elements of a symmetry group, as we exploit below and more systematically in \cite{uspawel}.

The second situation concerns mixing of the initial generator with generators of different penalties as we proceed along the unitary trajectory. This is obviously the `complicated' scenario, which we will term operator growth, as in \cite{SSD,operatorgrowth}. Quite interestingly and counterintuitively, we will see that in holographic dualities both types of growths seem to be to dual to each other.


\section{Quantum complexity and gravity}\label{secIV}

In this section, we apply the previous ideas to study specific aspects of quantum gravity, such as the behavior of computational costs under general coordinate transformations, their connection to quantum chaos, and their dynamics in SYK. We go from the simplest examples towards the more complex ones, so we start by analyzing the cost of symmetry transformations.

\subsection{The cost of symmetry}\label{secIV.I}

In this section, we study the cost of various symmetry transformations. First, notice that the geometric approach to complexity is basically equal to the geometric view of Lie groups in physics. In particular, the cost function of an infinitesimal transformation is a norm on the Lie algebra of the theory, while finite distances are obtained by composing infinitesimal ones. For concreteness, consider a quantum theory in which certain continuous symmetry group $G$ acts naturally in the Hilbert space and in the operator algebra. Natural `gates' in this system are symmetry transformations:
\begin{equation}
U_{\textrm{target}}=U(g_{N})\cdots U(g_{1})\;,
\end{equation}
where $g_{i}\in G$ and $U(g)$ is a representation of $G$ in the Hilbert space. To study continuous paths, we can increase to infinity the number of gates, while decreasing the strength of each unitary. In the continuous limit, a gate is given by an infinitesimal symmetry transformation, which can be expanded on the Lie algebra of the group $G$:
\begin{equation}
U(\theta^{a})\equiv 1+i\theta^{a}t_{a}=1+i\tilde{H}(\theta)\;,
\end{equation}
where the $t_{a}$ are the hermitian generators of the symmetry group. In this situation, the instantaneous Hamiltonian $\tilde{H}(\theta)$ is a particular element of the Lie algebra, and the cost function is a norm on the algebra. If one of the generators is the Hamiltonian of the system, we can explore time evolution, but in general, there will be other directions in the symmetry group to explore.

As we are going to see, there are two important advantages of using symmetry groups as the `gate' set. The first is that for symmetry group transformations (at least the ones we consider), the penalty functions will be fixed up to a global choice of units. In this sense, symmetry transformations belong to the simple class described above, where the cost functions are just given by expectations values in the appropriate state. The second advantage is that the group structure allows us to handle the computation of instantaneous Hamiltonians. In this article, we consider rotations, boosts and some general coordinate transformations important for black hole physics. In \cite{uspawel} we consider the Virasoro group.

Let's start with the simplest example, which is that of $SU(2)$. The generators are the three components of the angular momentum. If the theory is rotationally invariant, the penalty factors associated to each direction must be the same $p_{J_{x}}=p_{J_{y}}=p_{J_{z}}$. We can thus fix the global units so that the penalties are equal to $1$. The cost function simplifies to:
\begin{equation}\label{ssym}
F(\tilde{H}(\theta))=\sqrt{\langle\psi\vert \tilde{H}(\theta)\tilde{H}(\theta)\vert\psi\rangle}\;.
\end{equation}
For example, for a finite rotation of angle $\theta$ around the unit vector $\overrightarrow{n}$, we have $U(\theta)=e^{-i \theta \overrightarrow{\hat{J}}\cdot \overrightarrow{n}}$, and the cost grows as:
\begin{equation}\label{theta}
C(e^{-i \theta \overrightarrow{\hat{J}}\cdot \overrightarrow{n}})=J\theta\;,
\end{equation}
where
\begin{equation}
J=\sqrt{\langle (\overrightarrow{\hat{J}}\cdot \overrightarrow{n})^{2}\rangle}\;.
\end{equation}
Such rotations are geodesics, see \cite{uspawel} for a general proof and references therein. For the very same reason, the complexity of a translation is given by:
\begin{equation}\label{X}
C(e^{-i x\hat{p}})=p\,x\;,
\end{equation}
where $\hat{p}$ is the state momentum and $x$ is the traversed distance. Here as well, unitaries driven by constant momentum operators define minimal geodesics in the submanifold of the unitary group associated to the subgroup of translations. This is due to the abelian nature of the group, which implies that the complexity manifold is flat in those directions. This is beacuse all nested commutators that appear in the computation of the intantaneous Hamiltonian vanish. Therefore, the metric does not depend on the point chosen, labelled by $P^{\rho}$. It only depends on the instantaneous velocities $dP^{\rho}$. The manifold is thus diffeomorphic to flat space, and minimal geodesics are given by straight lines, i.e unitary trajectories driven by constant momentum operators (for example the Hamiltonian).

One interesting aspect of these observations is that looking at the complexity of Hamiltonian time evolution alone, it is very opaque what is the consequence of complexity minimization. On the other hand, already at the level of simple symmetry transformations like rotations or translations, we see that to minimize complexity we need to minimize the path lengths in the space-time manifold in which the symmetries are acting. In other words, particles moving through geodesics in space-time are those who minimize their associated computational costs (at least for geodesics defined by symmetry flows in the manifold).

\subsubsection{Rotations of the angular momentum}\label{secIV.I.I}

Let us slightly complicate the scenario and ask for the cost of the following rotation:
\begin{equation}\label{rotation}
e^{iJ_{x}(\theta)}=e^{iJ_{z}\theta}e^{iJ_{x}}e^{-iJ_{z}\theta}=e^{i(J_{x}\cos \theta-J_{y}\sin\theta)}\;.
\end{equation}
This would be the simplest analogue of Heisenberg time evolution. To compute the cost of~(\ref{rotation}), we need to find the instantaneous Hamiltonian $\tilde{H}(\theta)$:
\begin{equation}
\tilde{H}(J_{x,y}(\theta),\frac{dJ_{x,y}(\theta)}{d\theta})= i \textrm{ad}_{J_{x,y}(\theta)}^{-1}(e^{-i\textrm{ad}_{J_{x,y}(\theta)}}-\mathds{1})(\frac{J_{x,y}(\theta)}{d\theta})=\sum\limits_{j=0}^{\infty}\frac{(-i\textrm{ad}_{J_{x,y}(\theta)})^{j}}{(j+1)!}(\frac{dJ_{x,y}(\theta)}{d\theta})\;,
\end{equation}
where $\textrm{ad}_{J_{x,y}(\theta)}(\frac{dJ_{x,y}(\theta)}{d\theta})=[J_{x,y}(\theta),\frac{dJ_{x,y}(\theta)}{d\theta}]$. Given the group structure, the nested commutators oscillate between $-iJ_{z}$ and $-J_{y}$, so the previous expression can be easily resumed to:
\begin{equation}
\tilde{H}=\frac{dJ_{x,y}(\theta)}{d\theta}+J_{z}(\cos (1)-1)+J_{y}(\theta)(1-\sin (1))\;.
\end{equation}
Having this expression it is trivial to compute the evolution of the cost for any given state using~(\ref{ssym}). This example shows how group structures allow exact evaluations, and how one actually computes computational costs.

\subsubsection{Boosts}\label{secIV.I.II}

This simple analysys becomes more interesting for the Lorentz group. To the already considered angular momentum $\overrightarrow{J}$, linear momentum $\overrightarrow{P}$ and Hamiltonian $H$, we need to add the boost vector $\overrightarrow{K}$. Without lack of generality, consider a boost $K_{x}$ along the $x$ direction. The cost of
\begin{equation}
e^{-iK_{x}\eta}\;,
\end{equation}
is trivial and given by:
\begin{equation}
C(e^{-iK_{x}\eta})_{\psi}=\sqrt{\langle \psi\vert K_{x}\vert\psi \rangle^{2}}\eta\;.
\end{equation}
More interesting is the behavior of the relative cost associated to the boost of the linear momentum. For homogeneous Lorentz transformations, the linear momentum transforms as a vector:
\begin{equation}
(P')^{\rho}=U^{-1}(\Lambda)\,P^{\rho}\,U(\Lambda)=\Lambda_{\ \mu}^{\rho}P^{\mu}\;.
\end{equation}
Therefore, if the initial unitary is a displacement by $x_{\rho}$ in the position of the state we have:
\begin{equation}
e^{iK_{x}\eta} e^{iP^{\rho}x_{\rho}}e^{-iK_{x}\eta}=e^{iP^{\rho}(\eta)x_{\rho}}=e^{i\Lambda_{\ \mu}^{\rho}(\eta)P^{\mu}x_{\rho}}\;,
\end{equation}
where:
\begin{equation}
\Lambda (\eta)=\begin{pmatrix}
\cosh\eta & -\sinh\eta\\ -\sinh\eta &\cosh\eta
\end{pmatrix}\;.
\end{equation}
Now, since the group of translations is abelian, all nested commutators that appear in the computation of the instantaneous Hamiltonian vanish. We simply get:
\begin{equation}
\tilde{H}(\eta)=\frac{dP^{\rho}(\eta)}{d\eta}x_{\rho}=\frac{d\Lambda_{\ \mu}^{\rho} (\eta)}{d\eta}P^{\mu}x_{\rho}\;.
\end{equation}
Besides, since the instantaneous Hamiltonian is just a linear combination of operators with the same penalty, the cost reduces to the standard norm in the considered state. If such state has momentum $p^{\mu}$ we obtain:
\begin{equation}
F(\tilde{H}(\eta),\vert\Psi_{p}\rangle)=\sqrt{\langle\Psi_{p}\vert \frac{dP^{\rho}(\eta)x_{\rho}}{d\eta}\frac{dP^{\sigma}(\eta)x_{\sigma}}{d\eta}\vert\Psi_{p}\rangle}\;.
\end{equation}
We conclude that the behavior of the computational cost under Lorentz boosts is simply given by:
\begin{equation}\label{lorentz}
F(\tilde{H}(\eta),\vert\Psi_{p}\rangle)=\sqrt{(\frac{d\Lambda_{\ \mu}^{\rho} (\eta)}{d\eta}p^{\mu}x_{\rho})^{2}}\;.
\end{equation}
Using~(\ref{lorentz}), for a massless state with momentum $p^{\mu}_{1}=(p,-p,0,0)$, the associated costs to initial displacements $\Delta t$ and $\Delta x$ are:
\begin{equation}
\begin{pmatrix}
C(e^{iP^{t}(\eta)\Delta t})_{\Psi_{p_{1}}}\\ C(e^{iP^{x}(\eta)\Delta x})_{\Psi_{p_{1}}}
\end{pmatrix}=\int\limits_{0}^{\eta} d\eta'\begin{pmatrix}
\Delta t \,p\,e^{\eta'}\\ \Delta x \,p\,e^{\eta'}
\end{pmatrix}=\begin{pmatrix}
\Delta t \,p\,(e^{\eta'}-1)\\ \Delta x \,p\,(e^{\eta'}-1)
\end{pmatrix}\;,
\end{equation}
while for a massive state with momentum $p^{\mu}_{2}=(m,0,0,0)$ we have:
\begin{equation}
\begin{pmatrix}
C(e^{iP^{t}(\eta)\Delta t})_{\Psi_{p_{2}}}\\ C(e^{iP^{x}(\eta)\Delta x})_{\Psi_{p_{2}}}
\end{pmatrix}=\begin{pmatrix}
\Delta t \,m\,(\cosh\eta-1) \\  \Delta x \,m\, \sinh\eta
\end{pmatrix}\;.
\end{equation}
Finally, for a massive state with velocity $p^{\mu}_{v}=(p,-v,0,0)$ and large hyperbolic angle we have:
\begin{equation}
\begin{pmatrix}
C(e^{iP^{t}(\eta)\Delta t})_{\Psi_{p_{1}}}\\ C(e^{iP^{x}(\eta)\Delta x})_{\Psi_{p_{1}}}
\end{pmatrix}\xrightarrow[\eta\rightarrow \infty]{}\int\limits_{0}^{\eta} d\eta'\begin{pmatrix}
\Delta t\,\frac{p+v}{2}\,e^{\eta'}\\\Delta x\,\frac{p+v}{2}\,e^{\eta'}
\end{pmatrix}=\begin{pmatrix}
\Delta t\,\frac{p+v}{2}\,(e^{\eta'}-1)\\ \Delta x\,\frac{p+v}{2}\,(e^{\eta'}-1)
\end{pmatrix}\;.
\end{equation}
Notice hat the relativistic causality bound, stating that nothing can travel faster than the speed of light, has a precise inprint on the possible complexity growths. In particular, it bounds the prefactor of the exponential growth to be less than or equal the time component of the momentum multiplied by the initial displacement. Indeed, since the relativistic bound is $v\leq p$ we have $(p+v)/2\leq 1$ and:
\begin{equation}
C(e^{iP^{t}(\eta)\Delta t})\leq p\Delta t\, e^{\eta}\;.
\end{equation}

\subsection{Chaos and black holes}\label{secIV.II}

Building upon previous results, in this section we describe how computational costs are sensitive to the universal behavior of black holes. The first main observation is that, given the equivalence principle (not necessarily at the horizon), complexity has to grow exponentially with a rate controlled by the redshift factor. This implies that it grows with the maximal Lyapunov exponent derived in \cite{bound}. The second observation is that the exponential growth is a universal aspect which does not depend on details of the infalling particle, nor even on its infalling velocity. Details of the infalling velocity are encoded in the prefactor (which otherwise is still universal with respect to the nature of the particle). Letting the infalling velocity approach the speed of light suggests a bound on such prefactor.

To derive such aspects from a general standpoint, we consider the following $(d+2)$-dimensional geometry, which may admit a dual $(d+1)$-dimensional field theory formulation at finite temperature:
\begin{equation}\label{holob}
ds^{2} = F(\rho)\left(-h(\rho) dt^2 + d\ell^2 \,\right) + \frac{d\rho^2}{h(\rho)} \;.
\end{equation}
Here, $F(\rho)$ is the warp factor controlling the asymptotic behavior of the geometry at large $\rho$, and $h(\rho)$ models thermal effects. It has a simple zero at the horizon, $h(\rho_0)=0$,  and
approaches unity at large values of $\rho$. The Hawking temperature can be found by the usual Euclidean formalism to be;
\begin{equation}
T= \frac{h'_0}{4\pi} \sqrt{F_0}\;,
\end{equation}
where $F_0 \equiv F(\rho_0)$.

The blackening factor $h(\rho)$ is not universal, since it depends on the black hole considered. But as it is well known, it shows a definite universal structure near the horizon. This can be seen by taking the near horizon limit, in which $F(\rho)\rightarrow F_0$ and $h(\rho)\rightarrow h'_0\,(\rho-\rho_0)$ and the metric becomes:
\begin{equation}
ds^{2} = -F_0 \,h'_0\,(\rho-\rho_0)\,dt^2 + \frac{d\rho^2}{h'_0\,(\rho-\rho_0)}+ F_0 \,d\ell^2\;.
\end{equation}
The proper distance to the horizon is:
\begin{equation}
\rho_{p}=2\sqrt{\frac{\rho-\rho_0}{h'_0}}\;.
\end{equation}
Measuring radial distances with such coordinate, and using the relation for the Hawking temperature, the metric shows its well known universal character:
\begin{equation}
ds^{2} = -(2\pi T\rho_{p})^{2}\,dt^2 + d\rho_{p}^2+ds_{\perp}^{2}\equiv ds^{2}_{\textrm{univ}}+ds_{\perp}^{2} \;,
\end{equation}
where $ds^{2}_{\textrm{univ}}$ concerns the universal part, and $ds_{\perp}^{2}$ stands for the transversal coordinates. There is no real universality coming from the transverse metric, apart from the trivial flat space approximation for sufficiently small horizon patches. For the present pourposes, transverses directions play no role, since we will be considering radial geodesics for which $d\ell^2=0$. 

The universal behavior concerning the time and radial parts of the metric can be made more recognizable by defining the dimensionless time variable $\omega = 2\pi T t$, so that:
\begin{equation}
ds^{2} = -\rho_{p}^{2}\,d\omega^2 + d\rho_{p}^2+\cdots  \;,
\end{equation}
which is nothing but Rindler spacetime. This neatly shows that the near horizon region is just flat space in general relativity, and facilitates the coordinate transformation that takes us to the usual Minkoswki manifold. This is given by:
\begin{eqnarray}
&T=\rho\sinh\omega &\nonumber\\
&X=\rho\cosh\omega &\;,
\end{eqnarray}
in which the metric becomes;
\begin{equation}
ds^{2}_{\textrm{univ}}=-dT^{2}+dX^{2}
\end{equation}
Given the previous coordinate transformation, and defining the usual proper time variable as $d\tau= \rho_{p}\,d\omega$, the transformation between the momentum operators associated to each reference frame is given by:
\begin{equation}
\begin{pmatrix}
P_{\tau}\\ P_{\rho}
\end{pmatrix}=\begin{pmatrix}
\cosh\omega & -\sinh\omega\\ -\sinh\omega &\cosh\omega
\end{pmatrix}\begin{pmatrix}
P_{T}\\ P_{X}
\end{pmatrix}\equiv \Lambda_{\omega}\begin{pmatrix}
P_{T}\\ P_{X}
\end{pmatrix}\;.
\end{equation}
These relations just state that the transformation between the Mikowski frame to the Rindler frame is just a time dependent Lorentz boost.

Since the coordinate transformation is a Lorentz boost, the results of the previous section apply. As long as the equivalence principle holds, freelly falling trayectories will have constant momentum $p$ in the Minkowski frame, and therefore we conclude that:
\begin{equation}
F_{1}(\frac{dP^{\rho}_{\textrm{Rindler}}(\omega)x_{\rho}}{d\omega},\vert\Psi_{p}\rangle)=\sqrt{(\frac{d\Lambda_{\ \mu}^{\rho} (\omega)x_{\rho}}{d\omega}p^{\mu}_{\textrm{Minkowski}}x_{\rho})^{2}}\;.
\end{equation}
This implies that the cost of a massless infalling state with momentum $p^{\mu}_{1}=(p,-p,0,0)$, associated to initial displacements $\Delta \tau$ and $\Delta \rho$, is given by:
\begin{equation}
\begin{pmatrix}
C(e^{iP^{\tau}(\omega)\Delta \tau})_{\Psi_{p_{1}}}\\ C(e^{iP^{\rho}(\omega)\Delta \rho})_{\Psi_{p_{1}}}
\end{pmatrix}=\int\limits_{0}^{\omega} d\omega'\begin{pmatrix}
\Delta \tau\, p\,e^{\omega'}\\ \Delta \rho\,p\,e^{\omega'}
\end{pmatrix}=\begin{pmatrix}
\Delta \tau\, p\,(e^{\omega'}-1)\\ \Delta \rho\,p\,(e^{\omega'}-1)
\end{pmatrix}\;,
\end{equation}
while for a massive state with momentum $p^{\mu}_{2}=(m,0,0,0)$ we have:
\begin{equation}
\begin{pmatrix}
C(e^{iP^{\tau}(\omega)\Delta \tau})_{\Psi_{p_{2}}}\\ C(e^{iP^{\rho}(\omega)\Delta \rho})_{\Psi_{p_{2}}}
\end{pmatrix}=\begin{pmatrix}
\Delta \tau\,m\,(\cosh\omega-1) \\ \Delta \rho\, m\, \sinh\omega
\end{pmatrix}\;.
\end{equation}
Since $\omega=\frac{2\pi}{\beta}t$:
\begin{equation}
\begin{pmatrix}
C(e^{i P^{\tau}(t)\Delta \tau})_{\Psi_{p_{1}}}\\ C(e^{iP^{\rho}(t)\Delta \rho})_{\Psi_{p_{1}}}
\end{pmatrix}=\begin{pmatrix}
\Delta \tau\,p\,(e^{\frac{2\pi}{\beta}t}-1)\\ \Delta \rho\,p\,(e^{\frac{2\pi}{\beta}t}-1)
\end{pmatrix}\xrightarrow[t\gg\beta]{} \begin{pmatrix}
\Delta \tau\,p\,e^{\frac{2\pi}{\beta}t} \\  \Delta \rho\,p\, e^{\frac{2\pi}{\beta}t}
\end{pmatrix}\;,
\end{equation}
in the first scenario, while in the second:
\begin{equation}
\begin{pmatrix}
C(e^{iP^{\tau}(t)\Delta \tau})_{\Psi_{p_{2}}}\\ C(e^{i\,P^{\rho}(t)\Delta \rho})_{\Psi_{p_{2}}}
\end{pmatrix}=\begin{pmatrix}
\Delta \tau\,m\,(\cosh(\frac{2\pi}{\beta}t)-1) \\  \Delta \rho\,m\, \sinh(\frac{2\pi}{\beta}t)
\end{pmatrix}\xrightarrow[t\gg\beta]{}\begin{pmatrix}
\Delta \tau\,\frac{m}{2}\,e^{\frac{2\pi}{\beta}t} \\  \Delta \rho\,\frac{m}{2}\, e^{\frac{2\pi}{\beta}t}
\end{pmatrix}\;.
\end{equation}
For a general infalling state with momentum $p^{\mu}_{2}=(p,-v,0,0)$, we would obtain
\begin{equation}
\begin{pmatrix}
C(e^{iP^{t}(\tau)\Delta \tau})_{\Psi_{v}}\\ C(e^{iP^{\rho}(t)\Delta \rho})_{\Psi_{v}}
\end{pmatrix}\xrightarrow[t\gg\beta]{} \begin{pmatrix}
\Delta \tau\,\frac{p+v}{2}\,e^{\frac{2\pi}{\beta}t} \\  \Delta \rho\,\frac{p+v}{2}\, e^{\frac{2\pi}{\beta}t}
\end{pmatrix}\;,
\end{equation}
There are a couple of important observations we can draw from these results. The first is that relative computational costs are sensitive to the universal structure of black holes, as dictated by their near horizon regions. These computational costs are not $1/N$ effects, but $\mathcal{O}(1)$ features that neatly codify the universal structure, as we were seeking in the introduction. The second observation is that this result rests on the equivalence principle. If the momentum operators in a freelly falling frame are constant, as they should if the equivalence principle holds, then the costs associated with an outside observer grow with the maximal Lyapunov exponent.

The second observation is that the universal Lyapunov growth applies to all freely falling trajectories. It even applies to particles moving faster than the speed of light. In this sense, the Lyapunov growth might also apply to bulk theories with causality violations. On the other hand, the specifications of the infalling particle velocity neatly appear in the long-time asymptotics of the prefactor accompanying the exponential growth. This prefactor is still universal. It does not depend on the nature of the particle, just on its four-momentum (its infalling trajectory). This observation suggests a further bound on the growth of chaos for quantum theories having local gravity duals (at least as defined by complexity evolution). From the gravity perspective, the strongest growth is obtained by saturating causality at the local Minkowski level and letting the infalling particle move at the speed of light. Looking at the previous formulas, the results suggest that for theories with causal gravity duals we expect:
\begin{equation}\label{bh}
C\leqslant \Delta \tau\,E e^{\frac{2\pi}{\beta}t}\;,
\end{equation}
for the behavior of the complexity of the momentum operator associated to the infalling particle. We stress that the new part of the bound is in the prefactor and that $\Delta \tau$ is the initial displacement, which sets the initial perturbation.

It would be nice to have a clear dual of this growth, which is otherwise totally rooted in the growth of the radial momentum and the proper energy of the infalling particle, which are bounded by the previous relation without the initial displacement prefactor.  Recently, in \cite{Sfall} it has been proposed that such growths might be related to the size of the dual operator, as defined below when considering the cost growth of SYK. In the SYK scenario, we will see that indeed the cost growth is controlled by the operator size. The problem with the operator size is that it is a quantity specially built for spin systems, and not so clearly defined for QFT's. During the discussion of the penalty functions in CFT's~(\ref{secII.II}), we noticed that due to the operator product expansion, we do not need to include operator products. We just need to include local operators of all possible scaling dimensions. From this perspective, what grows under Heisenberg time evolution is the average scaling dimension of the perturbed operator, where we remind that the average scaling dimensions might be defined as~(\ref{aver}). We thus expect a duality between the growth of proper energy and the growth of the scaling dimensions. We remind that from this scaling dimension perspective, penalty factors just allow observing such scaling dimensions dynamics.

This proposal is interesting for various reasons. First, it is well known that there is a precise relation between energies in AdS and scaling dimensions in the boundary. This is valid for any space-time dimension. In other words, in the context of AdS/CFT, scaling dimensions gravitate. It is thus natural to relate the growth of proper energy and momentum of the infalling particle to the growth of the average scaling dimension of the dual operator. Besides, if the growth of the scaling dimension continues for a sufficiently long time, we will eventually need to account for its backreaction on the geometry. This would explain the expected backreaction of the infalling particle in the gravitational description, a feature that lies at the root of the behavior of out of time-ordered correlation functions \cite{bound}. The second interesting aspect is that, if such duality is correct, from the previously found behavior of proper energies and relation~(\ref{bh}), we expect an exponential growth for such average scaling dimensions and a universal behavior of the prefactor. More concretely we expect a bound of the type:
\begin{equation}\label{scalingbound}
\overline{\Delta}(t)\leqslant \Delta e^{\frac{2\pi}{\beta}t}\;,
\end{equation}
where $\Delta$ is the average scaling dimension of the perturbed operator. In the next section, when analyzing the cost growth in SYK, we will describe these features as well. At infinite temperature, the lower bound we are able to compute does not saturate the previous one, giving hope that it is indeed a non-trivial bound.

\subsection{The cost of operator growth in SYK}\label{secIV.III}
 
In section~(\ref{secII.III}) we explained how computational costs simplify whenever the initial operator does not mix with other operators, or whenever it just mixes with other operators of equal penalties. These were called `simple growths'. In the context of AdS/CFT \cite{adscft}, the black hole analysis we have performed would apply to the bulk description, in which the theory is weakly interacting and operators do not grow, in the sense of \cite{SSD,operatorgrowth}. But complexity does grow, and it does so in a very non-trivial exponential manner, as we just described. To try to understand this exponential complexity growth from a dual perspective, we can seek to compute the cost of Heinsenberg time evolution:
\begin{equation}
U(\mathcal{O},t)\equiv e^{iHt}e^{i\mathcal{O}}e^{-iHt}=e^{i\mathcal{O}(t)}\;,
\end{equation}
in the thermal state. This seems a challenging task. Since the dual theory is strongly coupled, the evolution of $\mathcal{O}(t)$ is not going to be simple at all, and the operator will mix with operators associated with different penalties. We thus need to take care of the penalties by using formula~(\ref{metricstate1}), or its CFT version~(\ref{metricstate2}).

Now, for generic theories, even knowing the dynamics of operator growth, the computation seems challenging. As explained better in section~(\ref{secII.t}), this is because once we have $\mathcal{O}(t)$ and $\frac{d\mathcal{O}(t)}{dt}$, we need to insert them in the expression for the instantaneous Hamiltonian~(\ref{HXY}), find all nested commutators, and add them up.

At the time being, this computation seems out of reach. We will content ourselves with evaluating a lower bound for the evolution of the computational cost in the case of SYK, using the recent results of Ref~\cite{operatorgrowth}. SYK models \cite{kitaev,remarks} are models of $N$ Majorana fermions interacting through random k-body interactions:
\begin{equation}\label{SYK2}
H= i^{q/2}\sum\limits_{1\leq i_{1}<\cdots<i_{q}\leq N}J_{i_{1}\cdots i_{q}}\chi_{i_{1}}\cdots \chi_{i_{q}}\;.
\end{equation}
Each term in the above sum contains $q$ Majorana fermions and the couplings are real random numbers with zero mean and variance equal to $\langle J_{i_{1}\cdots i_{q}}^{2}\rangle = J^{2}\frac{(q-1)!}{N^{q-1}}$.

Although the motivations to study these models seem very well known by this time, let us describe them briefly here for completeness. First, these models have an infrared conformal phase and were shown to have holographic duals and saturate the chaos bound by Kitaev \cite{kitaev}, see \cite{remarks} for a complete discussion. Second, this is a new class of solvable models in the large-N limit, intimately connected with the previously known tensor models \cite{witten, Klebanov}. Also, the zero temperature entropy reproduces black hole entropy, as shown in \cite{sachdev}. There are expectations that these models could potentially be created in the lab \cite{labSYK}. Finally, these models are excellent models for discussions of quantum chaos and thermalization \cite{kbody,kitaev,usfree,usfreeblack,usperm,reviewkbody,Sonner,Haque}, since dissipative phenomena can be treated analytically, and for the same reasons they can be used to extract generic conclusions on the behavior of entanglement dynamics in large-N theories \cite{usfreeblack,usperm,usdf}.

For the concerns of this article, SYK is also interesting because it is both a spin system and CFT, so it is the perfect setup to test possible generalizations of Nielsen approach to spin systems. In particular, in exact analogy to the case in which we have $N$ spins degrees of freedom, and any instantaneous Hamiltonian can be expanded in the basis of generalized Pauli matrices, in the present scenario we can expand any instantaneous Hamiltonian as:
\begin{equation}\label{expansion}
\tilde{H}=\sum\limits_{s}\sum\limits_{i_{1}<\cdots <i_{s}}c_{i_{1}\cdots i_{s}}\chi_{i_{1}}\cdots \chi_{i_{s}}\;.
\end{equation}
Hermiticity of $\tilde{H}$ implies that the coefficients are either real or pure imaginary, and in this case they can be easily obtained by defining the standard inner product:
\begin{equation}
(\mathcal{O},\mathcal{O})=\textrm{Tr}[\rho_{\textrm{mixed}}\mathcal{O}^{\dagger}\mathcal{O}]\;,
\end{equation}
where $\rho_{\textrm{mixed}}=\mathds{1}/2^{N/2}$ is the maximmally mixed density matrix in the Hilbert space of $N$ Majorana fermions. We have normalized the fermions so that $\chi^{2}=\mathds{1}$. Therefore:
\begin{equation}
(\tilde{H},\tilde{H})=\sum\limits_{s}\sum\limits_{i_{1}<\cdots <i_{s}}\vert c_{i_{1}\cdots i_{s}}\vert^{2}\;.
\end{equation}
Now notice that, on average, the SYK model is invariant with respect to a relabelling of the fermions. This implies that all operators of size $s$, i.e operators of the form $\chi_{i_{1}}\cdots \chi_{i_{s}}$, have the same average scaling dimension \footnote{To define the scaling dimension of a product of operators we can use the operator product expansion and then compute the average scaling dimension of the resulting combination.}. Equivalently, the scaling dimension is a function of the size of the operator $\overline{\Delta}= f(s)$. We conclude that in SYK, the penalties can be equivalently defined in terms of $s$ or $\overline{\Delta}$, giving strong support that in general CFT's, it is the scaling dimension the property that should be `punished', as put forward in section~(\ref{secII.II}).


Following the steps described in~(\ref{secII}), it is natural to define a projector into the space of equal penalty factors, defined there as the space of equal scaling dimension $\Delta$. In SYK, proyectors into the space of equal size operators are naturally organized by their average scaling dimension:
\begin{equation}
\hat{P}_{\Delta}(\tilde{H}(t))\xrightarrow[\textrm{SYK}]{}\hat{P}_{s}(\tilde{H}(t))=\sum\limits_{i_{1}<\cdots <i_{s}}c_{i_{1}\cdots i_{s}}\chi_{i_{1}}\cdots \chi_{i_{s}}\;.
\end{equation}
Notice that:
\begin{equation}
(\hat{P}_{\Delta (s)}(\tilde{H}(t)),\hat{P}_{\Delta (s)}(\tilde{H}(t)))=\sum\limits_{i_{1}\cdots i_{s}}\vert c_{i_{1}\cdots i_{s}}\vert^{2}\equiv \tilde{P}_{s}(t).
\end{equation}
Using~(\ref{metricstate2}), the cost of such Hamiltonian in the infinite temperature or maximally mixed state is:
\begin{equation}\label{metricSYK}
F_{1}(\tilde{H}(t),\rho_{\textrm{mixed}})=\sqrt{\sum\limits_{\Delta(s)}p_{\Delta (s)}^{2}\tilde{P}_{s}(t)}=\sqrt{\sum\limits_{s}p_{\Delta (s)}^{2}\tilde{P}_{s}(t)}\;.
\end{equation}
Now we consider perturbing the thermal state with a unitary matrix $V(t)=e^{i\chi_{1}}$. This is like setting the first fermion in a certain coherent state. As time evolves:
\begin{equation}
V(t)=e^{iHt}e^{i\chi_{1}}e^{-iHt}=e^{i\chi_{1}(t)}\;,
\end{equation}
where $\chi_{1}(t)=e^{iHt}\chi_{1}e^{-iHt}$ is the usual Heisenberg time evolution. Such operator can be expanded as:
\begin{equation}
\chi_{1}(t)=\sum\limits_{s}\sum\limits_{i_{1}<\cdots <i_{s}}c_{i_{1}\cdots i_{s}}(t)\chi_{i_{1}}\cdots \chi_{i_{s}}\;.
\end{equation}
This expansion was studied recently in \cite{operatorgrowth}. In the limit of large $q$, the following result was obtained:
\begin{eqnarray}\label{psyk}
P_{1}(t)&=&\vert c_{1}\vert^{2}=1-\frac{4}{q}\log \cosh \mathcal{J}t\nonumber\\
P_{s\neq 1}(t)&=& \sum\limits_{i_{1}<\cdots <i_{s}}\vert c_{i_{1}\cdots i_{s}}\vert^{2}=
\frac{2}{kq}\tanh^{2k} \mathcal{J}t\,\,\,\,\,\,\,\,\,\,\,\, s+1+(q-2)k\,\,\,\, k=1,2,3,\cdots \;.
\end{eqnarray}
To compute complexity, we need to extract the instananeous Hamiltonian driving the unitary at each differential amount of time. This is generically given by~(\ref{HXY}). Given the random nature of the dynamics, a lower bound on the growth can be found just by taking the first term, since the inclusion of all other terms will just increase the cost of the operator. The first term is the time derivative $d\chi_{1}(t)/dt$:
\begin{equation}
\frac{d\chi (t)}{dt}=\sum\limits_{i_{1}\cdots i_{s}}\frac{dc_{i_{1}\cdots i_{s}}(t)}{dt}\chi_{i_{1}}\cdots \chi_{i_{s}}\;.
\end{equation}
Using~(\ref{metricSYK}), the cost of such operator is:
\begin{equation}
F_{1}(\frac{d\chi (t)}{dt},\rho_{\textrm{mixed}})=\sqrt{\sum\limits_{s}p_{\Delta (s)}^{2}\tilde{P}_{s}(t)}\;,
\end{equation}
where we have defined:
\begin{equation}
\tilde{P}_{s}(t)\equiv \sum\limits_{i_{1}<\cdots <i_{s}}\vert \frac{dc_{i_{1}\cdots i_{s}}(t)}{dt}\vert^{2}\;.
\end{equation}
We need to relate $\tilde{P}_{s}(t)$ to the original $P_{s}(t)$. Since the phases of the coefficients in the expansion~(\ref{expansion}) are constant in time, the relation is as follows:
\begin{equation}
\tilde{P}_{s}(t)=(\frac{dP_{s}}{dt})^{2}\frac{1}{4P_{s}(t)}
\end{equation}
To finish the computation we just need to insert the penalties, perform the sum and integrate over time. We will explore a polynomial family of penalties, defined by:
\begin{equation}
p_{\Delta (s)}^{2}=\Delta^{r}\,\,\,\,\,\,\, r=1,2,3,\cdots
\end{equation}
Reminding that the scaling dimension of the fermions is $1/q$, in the large-N limit the average scaling dimension of $\chi_{i_{1}}\cdots \chi_{i_{s}}$ is $ \Delta_{\chi}=s/q$. Combining all details, we finally arrive at:
\begin{equation}\label{csyk}
C(e^{i\mathcal{O}(t\gg 1/\mathcal{J})})\geq c_{r}\frac{e^{r\mathcal{J}t}}{\sqrt{q}}=c_{r}\frac{e^{r\lambda_{L} t/2}}{\sqrt{q}}\;,
\end{equation}
where $c_{r}$ is a constant that depends on $r$ and that can be computed case by case. The first two cases are $c_{1}=1/\sqrt{2}$ and $c_{2}=\frac{1}{4}\sqrt{3/2}$. Also we have used the expression for the SYK Lyapunov exponent at infinite temperature $\lambda_{L}=2\mathcal{J}$.

To summarize, relation~(\ref{csyk}) is a lower bound on the computational cost growth of Heisenberg time evolution in SYK. Observe that all penalty choices, characterized by $r$, are sensitive to the chaos exponent. Qualitatively, at least in this case, the penalty choice does not affect the main feature. But we also observe that to match the expected chaos growth we should choose $p_{\Delta}=\Delta$. This result fits quite well with the arguments developed in~(\ref{secII.II.I}). For such penalty choice, the cost of the operator is a natural physical quantity to consider. It is just the average of the square scaling dimension.

Notice also that the average scaling dimension itself is just given by:
\begin{equation}\label{scalingSYK}
\overline{\Delta}(t)=\sum\limits_{s}\frac{s}{q}P_{s}(t)=\frac{\cosh (\lambda_{L}t)}{q}\rightarrow \frac{e^{\lambda_{L}t}}{2q}\;.
\end{equation}
Given the proposal of the last section, this should be dual to the growth of proper energy~(\ref{bh}). In this case, our proposal coincides with the proposal of \cite{Sfall}, but it is now understood as a very subtle example of the duality between energy and scaling dimensions in AdS/CFT.

Notice also that the growth~(\ref{scalingSYK}) does not saturate the bound~(\ref{scalingbound}), given the $1/2$ prefactor. Here $1/q$ would be the initial energy, corresponding to the scaling dimension of the initially perturbed fermionic degree of freedom. Of course, we are computing the growth at infinite temperature. It is possible that saturation occurs at low temperatures, where the Lyapunov growth also saturates to its maximal value. But at any rate, this suggests that the bound~(\ref{bh}) is not trivial since it is not saturated by default. It would be interesting if it is able to discriminate between theories with maximal Lyapunov growth but non-local gravity duals.

\subsubsection{Saturation to linear growth after the scrambling time}\label{secIV.III.I}

The complexity of the operator $e^{i\chi_{1}(t)}$ has been shown to be controlled by the growth of the operator $\chi_{1}(t)$. The consequence is that complexity grows exponentially fast, and it is controlled by the chaos exponent. But such growth cannot continue forever. Soon after the operator has reached a size of $\mathcal{O}(N)$, there is no more room to grow and the operator growth process must saturate. More concretely, notice that the expansion:
\begin{equation}
\chi_{1}(t)=\sum\limits_{i_{1}<\cdots <i_{s}}c_{i_{1}\cdots i_{s}}(t)\chi_{i_{1}}\cdots \chi_{i_{s}}\;,
\end{equation}
can be understood as defining a probability distribution:
\begin{equation}
P_{i_{1}\cdots i_{s}}(t)(t)=\vert c_{i_{1}\cdots i_{s}}(t)\vert^{2}\;.
\end{equation}
The reason is that if $\chi_{1}(0)=\chi_{1}$, then we have $\sum\limits_{i}P_{i}(t)=1$ for all times. Moreover, Heisenberg time evolution drives such distribution to the uniform one at times greater than the scrambling time \cite{operatorgrowth}. The intuition is that at long times we can approximate the operator by a random operator, in which the probability of individual basis element is just the inverse of the total number of them. This is in the same spirit as the usual explanation of quantum thermalization by means of random states, see for example \cite{Lloyd,Page,usrandom}, and indeed it can be understood in similar terms, as we explain in the next section.

This same intuition holds for $d\chi_{1}(t)/dt$. Denoting its exapnsion by:
\begin{equation}
\frac{d\chi_{1}(t)}{dt}=\sum\limits_{i_{1}<\cdots <i_{s}}\frac{d c_{i_{1}\cdots i_{s}}(t)}{dt}\chi_{i_{1}}\cdots \chi_{i_{s}}\;,
\end{equation}
we observe again:
\begin{equation}
\textrm{Tr}(\frac{d\chi_{1}(t)^{\dagger}}{dt}\frac{d\chi_{1}(t)}{dt})=\sum\limits_{i_{1}<\cdots <i_{s}}\vert \frac{d c_{i_{1}\cdots i_{s}}(t)}{dt} \vert^{2}=\textrm{constant}\;.
\end{equation}
For example, in SYK for large-q such constant is easily found to be $2J^{2}/q$. Since the sum of the squares is constant, the expansion coefficients of the derivative also behave as a probability distribution. More interestingly, this argument holds as well for the exact instantaneous Hamiltonian. The exact expression for the instantaneous Hamiltonian was:
\begin{equation}\label{exact}
\tilde{H}(t)=\tilde{H}(\mathcal{O}(t),\frac{d\mathcal{O}(t)}{dt})=i \textrm{ad}_{\mathcal{O}(t)}^{-1}(e^{-i\textrm{ad}_{\mathcal{O}(t)}}-\mathds{1}) (\frac{d\mathcal{O}(t)}{dt})=\sum\limits_{j=0}^{\infty}\frac{(-i\textrm{ad}_{\mathcal{O}(t)})^{j}}{(j+1)!}(\frac{d\mathcal{O}(t)}{dt})\;.
\end{equation}
Even if this is a complicated expression, we will always be able to write it in the complete basis:
\begin{equation}
\tilde{H}(t)=\sum\limits_{s}\sum\limits_{i_{1}<\cdots <i_{s}}c_{i_{1}\cdots i_{s}}^{\tilde{H}}(t)\chi_{i_{1}}\cdots \chi_{i_{s}}\;.
\end{equation}
The interesting obervation is that, given the exact form~(\ref{exact}), the following expression holds:
\begin{equation}\label{trace}
\textrm{Tr}(\tilde{H}^{\dagger}(t)\tilde{H}(t))=\sum\limits_{s}\sum\limits_{i_{1}<\cdots <i_{s}}\vert c_{i_{1}\cdots i_{s}}^{\tilde{H}}(t)\vert^{2}=\textrm{constant}\equiv H^{2}
\end{equation}
This is because such expression is valid term by term in~(\ref{exact}), since for general time evolved operators we have:
\begin{equation}
\textrm{Tr}([A(t),B(t)])=\textrm{Tr}(U^{-1}(t)[A(0),B(0)]U(t))=\textrm{Tr}([A(0),B(0)])\;.
\end{equation}
For the same reasons as for $\chi_{1} (t)$, we expect $d\chi_{1}(t)/dt$ and the instantaneous Hamiltonian to reach stationarity at long times. These time-scales are obviously of the same order as the time by wich the operator $\chi_{1} (t)$ itself reaches stationarity. For $\tilde{H}(t)$, this means that on average, at long times, all coefficients are equal to $H^{2}/\Omega$, where $\Omega\rightarrow\frac{2^{N}}{q}$ for large-N. The complexity growth at long times (longer than the scrambling time) is thus given by:
\begin{equation}
C(e^{i\chi_{1}(t)})\xrightarrow[t\gg \frac{1}{2\mathcal{J}}\log N]{}\int dt \sqrt{\frac{H^{2}}{\Omega}\sum\limits_{\Delta }\Delta (s)^{r}\binom{N}{s}}\propto t\;.
\end{equation}
To compute the proportionality factor, the only thing that needs to be found is $H$. Again, this is a difficult task but one that might be acutally achieved. This is beacuse to compute $H$, and therefore to compute the growth rate a long times, we do not need to go to long times. Since $H$ is constant we can compute it at any non-zero small time, and we expect simplifications, or that approximations can be made. We hope to report on this in the future.

\subsection{Long times, Lloyd's bound and bulk duals}\label{secIV.IV}

The exponential complexity growths derived for black holes and SYK (relations~(\ref{bh}) and~(\ref{csyk})), might lead to an inconsistency with Lloyd's bound \cite{lloyd}. In the geometric approach to quantum complexity,  Lloyd's bound simply tells that the maximal complexity growth is given by constant Hamiltonian evolution. This is simply:
\begin{equation}\label{lloyd}
C\leq M t\;,
\end{equation}
where $M$ is the mass of the black hole or the total energy of the system. For perturbations around equilibrium, the growth found in this article is:
\begin{equation}
C\propto  E  e^{\frac{2\pi}{\beta}t}\;,
\end{equation}
where $E$ is the energy of the perturbation and the proportionality factors depends on the initial conditions, see~(\ref{secIV.II}). Although the exponential growth is certainly fast, for small perturbations and for times smaller than the scrambling time, it is actually slower than the linear growth~(\ref{lloyd}). This is because there is a hierarchy between $M$ and $E$, given by $M\sim S E$, where $S$ is the entropy of the black hole. But if such exponential growth would continue forever, eventually it would bypass Lloyd's bound, leading us to a certain tension.

From a bulk description in AdS/CFT, it was shown in \cite{SStanford} that such growth does not continue forever. For times larger than the scrambling time, where one would begin to violate Lloyd's bound, we need to include the backreaction of the perturbation on the geometry. This implies a linear growth at long times, see also \cite{Snegative}. It is interesting that this sudden change of dynamics, as determined by general relativity, seems to be anchored in the previous argument, which states that~(\ref{lloyd}) is the true maximal complexity growth. 

Quite strikingly, such dynamical transition was explicitly seen in SYK. It is ultimately due to the saturation of the operator growth process, which takes us from exponential to linear growth in the evolution of complexity. From the dual theory point of view, it is essentially a finite size effect. If the entropy is finite, the operator cannot grow forever, and the transition to linear growth will occur at sufficiently long times. What it is interesting is that such finite size effect is fully captured by the classical dynamics of general relativity, which otherwise is expected to only capture semiclassical aspects. In this section, we want to argue that the ideas generalize to any dual theory.

For a generic theory, one would consider a perturbation at $t=0$ of the type $e^{i\mathcal{O}(0)}$, for some given observable $\mathcal{O}$. In QFT's this perturbation could be a smeared operator over some time slice. As time evolves, the operator mixes in a complicated a way, but at long times it will reach a simple stationary behavior, when proyected over some state, for example the thermal one. Since time evolution leaves the canonical density matrix invariant, it follows that:
\begin{eqnarray}\label{sums}
&\textrm{Tr}(\rho_{\beta}\mathcal{O}^{\dagger}(t)\mathcal{O}(t))=&\textrm{constant}\equiv \mathcal{O}^{2}\\
&\textrm{Tr}(\rho_{\beta}\frac{d\mathcal{O}^{\dagger}(t)}{dt}\frac{d\mathcal{O}(t)}{dt})=&\textrm{constant}\equiv (d\mathcal{O}/dt)^{2}\nonumber\\
&\textrm{Tr}(\rho_{\beta}\tilde{H}(t)^{\dagger}\tilde{H}(t))=&\textrm{constant}\equiv\tilde{H}^{2}\;,
\end{eqnarray}
where we remind that the instantaneous Hamiltonian $\tilde{H}(t)$ is given by~(\ref{exact}). Now, for the same reasons that at long times unitary evolution drives quantum states to random states \cite{Lloyd,Page,usrandom}, unitary evolution will drive the operators $\mathcal{O}(t)$, $d \mathcal{O}(t)/dt$ and $\tilde{H}(t)$ to certain random operators, characterized by the fact that the modulus of the expansion coefficients are constant on average.  Therefore, for times larger than the scrambling time, complexity always grows linearly with time. Besides, at stationarity we expect such constant coefficients to be proportional to their associated probabilities in the thermal ensemble. In this situation, the linear growth is given by:
\begin{equation}
C(e^{i\mathcal{O}(t)})\xrightarrow[t\gg \beta\log N]{} t \sqrt{\tilde{H}^{2}\sum\limits_{\Delta }\Delta ^{r}\frac{e^{-\beta\Delta}}{Z}}\;,
\end{equation}
where $Z=\sum\limits_{\Delta}e^{-\beta\Delta}$. Again, we remark that $\tilde{H}^{2}$ can be computed at any small time scale, while at the same it controls the long time asymptotic growth of the computational cost.

\section{Conclusions}\label{secV}

It is still not fully understood how space-time distances in the bulk description of holographic dualities are to be represented in the CFT side. While bulk space-time distances are still mysterious in such sense, distances in the Hilbert space, or distances in the manifold of unitaries, have to be respected across dualities. They are just \emph{the same} if the equivalence of the underlying Hilbert space and of the microscopic Hamiltonians holds.

In this article, inspired by the geometric approach to quantum complexity developed by Nielsen and collaborators \cite{Nielsen1,Nielsen2,Nielsen3}, and by the recent ideas which relate gravity and complexity \cite{SusskindQC,Aaronson}, we have explored certain distance notions in the manifold of unitaries and in the Hilbert space. These notions have been defined in~(\ref{secII}) for generic CFT's and generic states. The definitions reduce to the ones given in  \cite{Nielsen1,Nielsen2,Nielsen3} whenever it makes sense to consider maximally mixed density matrices associated to finite dimensional Hilbert spaces. For CFT's, the definition~(\ref{metricstate2}) is based on the statement that the role played by the weight in spin systems is played by the scaling dimension in CFT's. This statement suggests by itself a natural interpretation of the penalty functions. They allow studying the average scaling dimensions of the appropriate operator. Also, the state dependence of formulas~(\ref{metricstate2}) turns out to be crucial to connect to gravity in CFT's, as shown in \cite{uspawel}.

After defining the framework, one of the most important observations of this work is to notice that the difficulties that appear in actual complexity computations (described in~(\ref{secII.II}) and~(\ref{secII.III})), disappear when considering unitary trajectories driven by generators of a certain symmetry group. As explicit examples, we computed the costs associated with rotations, translations, and boosts. In particular, we have shown that the cost associated to a boost of the momentum operator grows exponentially with the rapidity, and the prefactor carries information about the detailed space-time trajectory. Since the relation between the freely falling frame in black hole spacetimes and the static outside frame is a time-dependent Lorentz boost, we concluded that the cost associated to the evolution of momentum of a freely falling particle increases exponentially with the maximal Lyapunov exponent. The prefactor accompanying the exponential is still universal. It does not depend on the nature of the infalling particle, just on its infalling trajectory. This suggests a further bound on the growth of chaos (at least as defined by complexity evolution), which is obtained when we allow the infalling particle to saturate bulk causality. Therefore, at least for theories with a local gravitational dual near the horizon, we concluded that the evolution of complexity is bounded by~(\ref{bh}). Such prefactor might potentially be able to discriminate between theories with maximal Lyapunov growth that violate bulk causality and those that do not violate it. Given the provided definition of complexity for CFT's~(\ref{metricstate2}), this feature should translate into a bound on the growth of the average scaling dimension of the perturbed operator.

In the last section, we attempted to compute these type of distances in dual formulations. We have partially succeeded in SYK, where we were able to provide a lower bound on the cost growth. This lower bound nicely shows a Lyapunov growth and its dynamics is directly related to the dynamics of operator growth of the perturbed operator. Besides, the average scaling dimension does not saturate the new bound alluded before. From this lack of saturation, we cannot conclude that SYK fails to reproduce black holes physics since this computation is in the high energy limit of SYK. But it gives a partial hope that the new bound is non-trivial and that it might convert into a finer way of discriminating between theories with maximal Lyapunov growth, as alluded above.

Lastly, we have described the late type asymptotics of the cost growth. After the scrambling time, the perturbed operator stops growing due to finite size effects in a thermal ensemble and the process reaches stationarity. At such long times, we can approximate the operator by a random operator. Such saturation has a definite imprint in the complexity growth, turning the Lyapunov exponential growth into a linear growth in time, where the slope of the linear growth can be computed at small times. This avoids a hypothetical tension between Lloyd's bound and the exponential growth, and also nicely corresponds to the gravitational dynamics derived in \cite{SStanford}.

\section*{Acknowledgements}

We want to thank Jose Barb\'on, Pawel Caputa and Horacio Casini for useful discussions. We are also grateful to the Yukawa Institute for Theoretical Physics for hospitality. This work was supported by the Simons foundation through the It From Qubit Simons collaboration.





\newpage
\begin{thebibliography}{00}

 \bibitem{adscft}
  J.~M.~Maldacena,
  {\em The Large N limit of superconformal field theories and supergravity},
  Adv.\ Theor.\ Math.\ Phys {\bf 2} (1998),
  [arXiv:9711200 [hep-th]].

 \bibitem{Harlow}
D.~Harlow,
  {\em TASI Lectures on the Emergence of the Bulk in AdS/CFT},
  [arXiv:1802.01040 [hep-th]].

\bibitem{fieldopcorrespondence}
    Gubser, S. S. and Klebanov, Igor R. and Polyakov, Alexander M,
    {\em Gauge theory correlators from noncritical stringctheory},
    Phys. Lett. {\bf 428}, 105-114 (1998),
  [arXiv:9802109 [hep-th]],  
  
    W.~Edward,
     {\em Anti de Sitter space and holography},
    Adv. Theor. Math. Phys. {\bf 2}, 253-291 (1998),
  [arXiv:9802150 [hep-th]].

\bibitem{Takayanagi}
   S.~Ryu and T.~Takayanagi,
   {\em Holographic derivation of entanglement entropy from the anti-de Sitter/Conformal field theory correspondence},
   Phys. \ Rev.\ Lett.\ {\bf 96} (2006) 181602,
[arXiv:0603001 [hep-th]].

\bibitem{LM}
  A.~Lewkowycz and J.~Maldacena,
{\em Generalized gravitational entropy},
 JHEP {\bf 1308} (2013) 090,
 [arXiv:1304.4926 [hep-th]].

 \bibitem{SS}
 S.~Shenker and D.~Stanford,
  {\em Black holes and the butterfly effect},
  JHEP {\bf 1403} (2014) 067,
  [arXiv:1306.0622 [hep-th]].
  
  S.~Shenker and D.~Stanford,
  {\em Stringy effects in scrambling},
  JHEP {\bf 1505} (2015) 132,
  [arXiv:1412.6087 [hep-th]].
  
  \bibitem{bound}
J.~Maldacena, S.~H,~Shenker and D.~Stanford,
{\em A bound on chaos},
JHEP {\bf 08} (2016) 106,
  [arXiv:1503.01409 [hep-th]].
  
    \bibitem{Sbook}
L.~Susskind and J.~Lindesay,
{\em  An Introduction To Black Holes, Information And The String Theory Revolution: The Holographic Universe},
Hackensack, USA: World Scientific (2005) 183.
  
   \bibitem{BM}
  J.~.L.~F.~Barbon and J.~M.~Magan,
  {\em Chaotic Fast Scrambling At Black Holes},
  Phys.\ Rev.\ D.\  {\bf 84} (2011) 106012,
  [arXiv:1105.2581 [hep-th]].

  
  J.~.L.~F.~Barbon and J.~M.~Magan,
  {\em Fast Scramblers, Horizons and Expander Graphs},
  JHEP {\bf 1208} (2012) 016,
  [arXiv:1204.6435 [hep-th]].

 \bibitem{Sfall}
L.~Susskind,
{\em  Why do Things Fall?},
 [arXiv:1802.01198 [hep-th]].
 
  \bibitem{Nielsen1}
  
 M.~A.~Nielsen,
{\em A geometric approach to quantum lower bounds},
 [arXiv:0502070 [quant-ph]].
  
   \bibitem{Nielsen2}
  
 M.~A.~Nielsen, M.~R.~Dowling, M.~Gu and A.~C.~Doherty
{\em Quantum computation as geometry},
Science {\bf 311}, 1133 (2006),
 [arXiv:0603161 [quant-ph]].
 
  
 \bibitem{Nielsen3}
  
 M.~R.~Dowling and M.~A.~Nielsen,
{\em The geometry of quantum computation},
 [arXiv:0701004 [quant-ph]].
 
  \bibitem{SusskindQC}
L.~Susskind,
  {\em Computational Complexity and Black Hole Horizons},
  Fortsch. Phys. {\bf 64}, 44-48 (2016),
  [arXiv:1402.5674 [hep-th]].
  
   \bibitem{Aaronson}
S.~Aaronson,
  {\em The Complexity of Quantum States and Transformations: From Quantum Money to Black Holes},
  [arXiv:1607.05256 [hep-th]].
 
 \bibitem{uspawel}
P.~Caputa and J.~Magan,
  {\em To appear}.
  
 
  \bibitem{Snegative}
   A.~Brown, L.~Susskind, B~Swingle, Y.~Zhao,
   {\em Quantum Complexity and Negative Curvature},
   Phys. \ Rev.\ D.\ {\bf 95} (2016) 045010.
  [arXiv:1608.02612 [hep-th]].
 
 
  \bibitem{operatorgrowth}
  
D.~Roberts, D.~Stanford and A.~Streicher,
{\em Operator growth in the SYK model},
 [arXiv:1802.02633 [hep-th]]. 
 
 


  
   \bibitem{SStanford}
D.~Stanford and L.~Susskind,
  {\em Complexity and Shock Wave Geometries},
  Phys.~Rev.~D {\bf 90}, 12 (2014),
  [arXiv:1406.2678 [hep-th]].
  
  \bibitem{lloyd}
S.~Lloyd,
  {\em Ultimate physical limits to computation},
  Nature {\bf 406}, 1047-1054 (2000),
  [arXiv:1102.0440 [hep-th]].
  
  \bibitem{JLMS}
D.~L.~Jafferis, A.~Lewkowycz, J.~Maldacena and S.~J.~Suh
  {\em Relative entropy equals bulk relative entropy},
  JHEP {\bf 06}, 004 (2016),
  [arXiv:1512.06431 [hep-th]].
  
 
 
  
 \bibitem{Myers}
R.~Jefferson and R.~Myers,
  {\em Circuit complexity in quantum field theory},
  JHEP {\bf 10}, (2017),
  [arXiv:1707.08570 [hep-th]].   
  
   \bibitem{Chapman}
S.~Chapman, M.~Heller, H.~Marrochio and F.~Pastawski,
  {\em Towards Complexity for Quantum Field Theory States},
  [arXiv:1707.08582 [hep-th]].    
  
   \bibitem{Yang}
R.~Yang,
  {\em A Complexity for Quantum Field Theory States and Application in Thermofield Double States},
  Phys.~Rev.~D {\bf 97}, 066004, (2018),
  [arXiv:1709.00921 [hep-th]].  
  
   \bibitem{PawelC}
P.~Caputa, N.~Kundu, M.~Miyaji, T.~Takayanagi and K.~Watanabe,
  {\em Anti-de Sitter Space from Optimization of Path Integrals in Conformal Field Theories},
  Phys.~Rev.~Lett {\bf 119}, 071602, (2017),
  [arXiv:1703.00456 [hep-th]].  
  
  P.~Caputa, N.~Kundu, M.~Miyaji, T.~Takayanagi and K.~Watanabe
   {\em Liouville Action as Path-Integral Complexity: From Continuous Tensor Networks to AdS/CFT},
  JHEP {\bf 11}, 097, (2017),
  [arXiv:1706.07056 [hep-th]].  
  
  \bibitem{Norihiro}
  K.~Hashimoto, N.~Iizuka, S.~Sugishita  
  {\em Thoughts on Holographic Complexity and its Basis-dependence},
 [arXiv:1805.04226 [hep-th]].  

  

   \bibitem{SSD}
D.~Roberts, D.~Stanford and L.~Susskind
  {\em Localized shocks},
  JHEP {\bf 03}, 051, (2015),
  [arXiv:1409.8180 [hep-th]].  
  
 

  
   \bibitem{kitaev}
  A.~Kitaev,
  {\em A simple model of quantum holography},
  Talks at KITP, April 7, 2015 and May 27, 2015.

   S.~Sachdev and J.~w.~Ye,
   {\em Gapless spin fluid ground state in a random, quantum Heisenberg ferromagnet},
   Phys. \ Rev.\ Lett.\ {\bf 70} (1993) 3339,
   arXiv:cond-mat/9212030 [cond-mat]. 
   
   
    \bibitem{remarks}
J.~Maldacena and D.~Stanford,
{\em Remarks on the Sachdev-Ye-Kitaev model},
Phys.\ Rev. \ D.\ {\bf 94} (2016) 106002.

    \bibitem{witten}
E.~Witten,
{\em An SYK-like model without disorder},  
   [arXiv:1610.09758 [hep-th]].
 
  
  \bibitem{Klebanov}
I.~R.~Klebanov, G.~ Tarnopolsky,
{\em Uncolored random tensors, melon diagrams, and the SYK models},  
   [arXiv:1611.08915 [hep-th]].


 \bibitem{sachdev}
 S.~Sachdev,
  {\em Bekenstein-Hawking Entropy and Strange Metals},
  Phys.\ Rev.\ X {\bf 5} (2015) 041025,
  [arXiv:1506.05111 [hep-th]].



   \bibitem{labSYK}
  
 I.~Danshita, M.~Hanada and M.~Tezuka,
{\em Creating and probing the Sachdev-Ye-Kitaev model with ultracold gases: Towards experimental studies of quantum gravity},
 [arXiv:1606.02454 [cond-mat]].


 \bibitem{kbody}
   L.~Benet and H.~A.~Weidenmueller,
  {\em Review of the k-body embedded ensebmles of gaussian random matrices},
 J.\ Phys.\ A {\bf 36} (2003) 3569,
  [arXiv:0207656 [cond-mat]].


\bibitem{usfree}
  J.~M.~Magan,
  {\em Random free fermions: An analytical example of eigenstate thermalization},
  Phys.\ Rev.\ Lett {\bf 116} (2016) 030401, 
  [arXiv:1508.05339 [quant-ph]].
  
    \bibitem{usfreeblack}
  
 J.~M.~Magan,
{\em Black holes as random particles: entanglement evolution in infinite range and matrix models},
 JHEP {\bf 1608} (2016) 081,
 [arXiv:1601.04663 [hep-th]].
  
   \bibitem{usperm}
  J.~M.~Magan,
  {\em K-local microscopic diffusion at the Sachdev-Ye-Kitaev model}, 
  [arXiv:1612.06765 [hep-th]].
  
   \bibitem{reviewkbody}
  V.~K.~Kota and N.~D.~Chavda,
  {\em Embedded random matrix ensembles from nuclear structures and their recent applications}, 
  .Int. J. Mod. Phys. {\bf E 27}, 1830001 (2018).
  
  
   \bibitem{Sonner}
  
 J.~Sonner and M.~Vielma,
{\em Eigenstate thermalization in the Sachdev-Ye-Kitaev model},
 JHEP {\bf 11} (2017) 149,
 [arXiv:1707.08013 [hep-th]].
 
 
  \bibitem{Haque}
  
 M.~Haque and P.~McClarty,
{\em Eigenstate Thermalization Scaling in Majorana Clusters: from Chaotic to Integrable Sachdev-Ye-Kitaev Models},
 [arXiv:1711.02360 [hep-th]].
 
 
  \bibitem{usdf}
  
  J.~M.~Magan,
 {\em De Finetti theorems and entanglement in large-N theories and gravity},
  Phys.\ Rev.\ D {\bf 96} (2017) 086002,
  [arXiv:1705.03048 [hep-th]].
  
 
  \bibitem{Lloyd}
S.~Lloyd and H.~Pagels,
{\em Complexity as thermodynamic depth},
Ann. Phys. {\bf 188} (1988) 186.

\bibitem{Page}
  D.~N.~Page,
  {\em Average entropy of a subsystem},
  Phys.\ Rev.\ Lett.\  {\bf 71} (1993) 1291,
  [arXiv:gr-qc/9305007].

\bibitem{usrandom}
  J.~M.~Magan and S.~Vandoren,
  {\em Entanglement in Fock space of random QFT states},
  JHEP {\bf 1507} (2015) 150,
  [arXiv:1504.01346 [hep-th]].








%
%
%
%
%
%
%
%
%
%
%
%
%
%
%
%
%
%
%
  


   




 
 
 
 
  \end{thebibliography}
\end{document}